\documentclass[prl,aps,epsf,showpacs,twocolumn]{revtex4}
\usepackage{times}
\usepackage{graphicx}
\usepackage{float}
\usepackage{latexsym,amsmath,amssymb,bm,euscript}
\usepackage{color}
\usepackage{subfigure}
\usepackage{epstopdf}
\usepackage[colorlinks=true,linkcolor=blue,citecolor=blue]{hyperref}
\usepackage{hyperref}

\usepackage{appendix}

\begin{document}

\title{Plaquette Ordered Phase and Quantum Phase Diagram in the Spin-$\frac{1}{2}$ $J_1$-$J_2$ Square Heisenberg Model}
\author{Shou-Shu Gong$^{1}$, Wei Zhu$^{1}$, D. N. Sheng$^{1}$, Olexei I. Motrunich$^{2}$, Matthew P. A. Fisher$^{3}$}
\affiliation{$^{1}$Department of Physics and Astronomy, California State University, Northridge, California 91330, USA\\
$^{2}$Department of Physics, California Institute of Technology, Pasadena, California 91125, USA\\
$^{3}$Department of Physics, University of California, Santa Barbara, California 93106-9530, USA}

\begin{abstract}
We study the spin-$\frac{1}{2}$ Heisenberg model on the square lattice with first- and second-neighbor antiferromagnetic interactions $J_1$ and $J_2$, which possesses a nonmagnetic region that has been debated for many years and might realize the interesting $Z_2$ spin liquid.
We use the density matrix renormalization group approach with explicit implementation of $SU(2)$ spin rotation symmetry and study the model accurately on open cylinders with different boundary conditions.
With increasing $J_2$, we find a N\'{e}el phase and a plaquette valence-bond (PVB) phase with a finite spin gap.
From the finite-size scaling of the magnetic order parameter, we estimate that the N\'{e}el order vanishes at $J_2/J_1\simeq 0.44$.
For $0.5 < J_2/J_1 < 0.61$, we find dimer correlations and PVB textures whose decay lengths grow strongly with increasing system width, consistent with a long-range PVB order in the two-dimensional limit.  The dimer-dimer correlations reveal the $s$-wave character of the PVB order.
For $0.44 < J_2/J_1 < 0.5$, spin order, dimer order, and spin gap are small on finite-size systems, which is consistent with a near-critical behavior.
The critical exponents obtained from the finite-size spin and dimer correlations could be compatible with the deconfined criticality in this small region.
We compare and contrast our results with earlier numerical studies.
\end{abstract}

\maketitle

\textit{Introduction.---}Quantum spin liquid (SL) is an exotic state of matter where a spin system does not form magnetically ordered state or break lattice symmetries even at zero temperature\cite{Nature_464_199}.
Understanding spin liquids is important in frustrated magnetic systems and may also hold clues to understanding the non-Fermi liquid of doped Mott materials and high-$T_c$ superconductivity\cite{RMP_78_17}.
While the exciting properties of SL such as deconfined quasiparticles and fractional statistics have been revealed in many artificially constructed systems \cite{PRL_86_1881, PRB_64_064422, PRB_66_205104,PRB_65_224412, PRL_94_146805, PRL_97_207204, NP_7_772, 1309_5669,AP_321_2,PRL_108_247206}, the possibility of finding spin liquids in realistic Heisenberg models has attracted much attention over the past 20 years due to its close relation to experimental materials.
The prominent example is the kagome antiferromagnet, where recent density matrix renormalization group (DMRG) studies
point to a gapped $Z_2$ SL\cite{PRL_101_117203,Science_332_1173,PRL_109_067201,NP_8_902, 1309_5669}
characterized by a $Z_2$ topological order and fractionalized spinon and vison excitations\cite{PRL_66_1773,PRB_44_2664, PRB_40_7387, PRB_60_1654,PRB_62_7850}.

One of the candidate models for SL is the spin-$\frac{1}{2}$ $J_1$-$J_2$ square Heisenberg model (SHM) with the Hamiltonian
\begin{equation}
H = J_1 \sum_{\langle i,j\rangle}S_i\cdot S_j + J_2 \sum_{\langle\langle i,j\rangle\rangle}S_i \cdot S_j ~,
\end{equation}
where the sums $\langle i,j\rangle$ and $\langle\langle i,j\rangle\rangle$ run over all the nearest-neighbor (NN) and the next-nearest-neighbor bonds, respectively. We set $J_1=1$.
The frustrating $J_2$ couplings suppress the N\'{e}el order and induce a nonmagnetic region around the strongest frustration point $J_2=0.5$\cite{PRB_38_9335,IJMP_02_203,PRB_41_9323,PRB_44_12050,PRB_54_9007,PRL_78_2216,PRB_60_7278,PRL_84_3173,PRL_87_097201,PRL_91_067201,PRL_91_197202,PRB_74_144422,PRB_78_214415,PRB_79_024409,EPJB_73_117,PRB_81_144410,PRB_85_094407,1112_3331,PRB_86_024424,PRB_79_224431,PRB_86_045115,PRB_86_075111,PRL_111_037202,PRB_88_060402,1309.6490,1308.2759}.
Different candidate states have been proposed based on approximate methods or small-size exact diagonalization calculations, such as plaquette valence-bond (PVB) state\cite{PRB_54_9007,PRL_84_3173,PRL_91_197202,PRB_74_144422,PRB_79_024409,PRB_85_094407,1309.6490}, the columnar valence-bond (CVB) state\cite{PRB_41_9323,PRB_44_12050,PRB_60_7278}, or a gapless SL\cite{PRL_87_097201,PRL_91_067201,PRL_111_037202,PRB_88_060402}.  However, the true nature of the nonmagnetic phase remains unresolved.

\begin{figure}[tbp]
\includegraphics[width = 1.0\linewidth,clip]{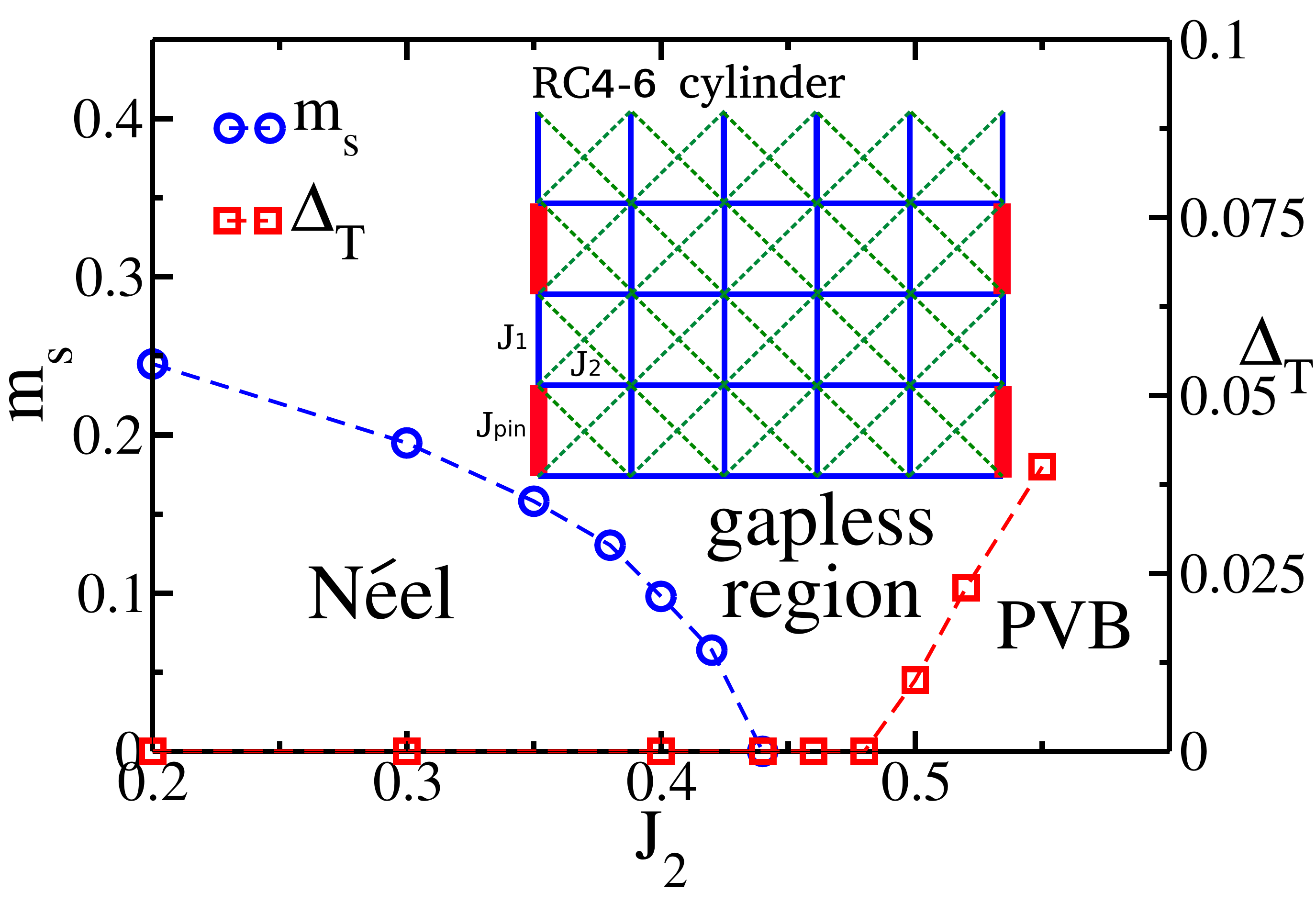}
\caption{(color online) Phase diagram of spin-$\frac{1}{2}$ $J_1$-$J_2$ SHM obtained by our $SU(2)$ DMRG studies. With growing $J_2$, the model has a N\'{e}el phase for $J_2 < 0.44$ and a PVB phase for $0.5 < J_2 < 0.61$.  Between these two phases, the finite-size magnetization and spin gap appear small in our calculations, consistent with a near-critical behavior. The main panel shows N\'{e}el order parameter $m_s$ and spin gap $\Delta_{\rm T}$ in the thermodynamic limit. The inset is a sketch of a RC4-6 cylinder; $J_{\rm pin}$ shows the modified odd vertical bonds providing the boundary pinning for dimer orders.}\label{phase_diagram}
\end{figure}

Recent DMRG study of the $J_1$-$J_2$ SHM \cite{PRB_86_024424} proposed a gapped $Z_2$ SL for $0.41 \leq J_2 \leq 0.62$ by establishing the absence of the magnetic and dimer orders, and by measuring a positive topological entanglement entropy term close to the value $\ln 2$ expected for a $Z_2$ SL\cite{PRL_96_110404, PRL_96_110405}.
Very recent variational Monte Carlo (VMC) work\cite{PRB_88_060402} proposed a gapless $Z_2$ SL for $0.45 \lesssim J_2 \lesssim 0.6$.
On the other hand, recent DMRG studies\cite{PRL_110_127203,PRL_110_127205,arxiv_1306_6067} of another bipartite frustrated system---the $J_1$-$J_2$ spin-$1/2$ honeycomb Heisenberg model---found a PVB phase in the nonmagnetic region, with a possible SL phase between the N\'{e}el and PVB phases\cite{arxiv_1306_6067} or with a direct N\'{e}el to PVB transition characterized by deconfined quantum criticality \cite{PRL_110_127203,PRL_110_127205, arxiv_1306_6067, PRB_70_144407,Science_303_1490}.
These studies\cite{PRL_110_127205,arxiv_1306_6067} 
also found that in the nonmagnetic region the convergence of DMRG in wider systems, which is controlled by the number of states kept, is crucial for determining the true nature of the ground state.

In this Letter, we reexamine the nonmagnetic region of the $J_1$-$J_2$ SHM using DMRG with $SU(2)$ spin rotation symmetry\cite{SU2}. We obtain accurate results on wide cylinders
by keeping as many as $36000$ $U(1)$-equivalent states. We find a N\'{e}el phase below $J_2 \simeq 0.44$ and a nonmagnetic region for $0.44 < J_2 < 0.61$ by finite-size scaling of the magnetic order parameter. In the nonmagnetic region, we establish a PVB order for $J_2 > 0.5$---in contrast to the previous proposal of a gapped $Z_2$ SL\cite{PRB_86_024424}---by observing that the PVB decay length grows strongly with increasing system width.
We identify the PVB order as the $s$-wave plaquette\cite{PRB_74_144422} by studying dimer-dimer correlations.
For $0.44 < J_2 < 0.5$, we find that the magnetic order, valence-bond crystal (VBC) orders, and spin excitation gap are small on finite-size systems, suggesting a near-critical behavior.  The magnetic and dimer critical exponents at $J_2 = 0.5$ are roughly similar to the values found for the deconfined criticality in the $J$-$Q$ models on the square and honeycomb lattices\cite{PRL_98_227202, PRL_100_017203, JSM_2008_02009, PRB_80_180414, PRB_80_212406, PRL_104_177201,PRL_111_137202, PRL_111_087203},
which is consistent with the deconfined criticality scenario conjectured also for the $J_1$-$J_2$ model in Ref.~\cite{PRB_85_134407}.

We establish the phases based on high accuracy DMRG results on cylinders\cite{converge}.
The first cylinder is the rectangular cylinder (RC) with closed boundary in the $y$ direction and open
boundaries in the $x$ direction.
We denote it as RC$L_y$-$L_x$, where $L_y$ and $L_x$ are the number of sites in the $y$ and $x$ directions;
the width of the cylinder is $W_y = L_y$ (see the inset of Fig.~\ref{phase_diagram}).
To study the dimers oriented in the $y$ direction, we can induce such an order near the open boundaries by modifying every other NN vertical bond on the boundary to be $J_{\rm pin} \neq J_1$ as illustrated in Fig.~\ref{phase_diagram}.
The second geometry is the tilted cylinder (TC), as shown in Fig.~\ref{dimer_TC}(a), when discussing VBC order.

\begin{figure}[tbp]
\includegraphics[width = 1.0\linewidth,clip]{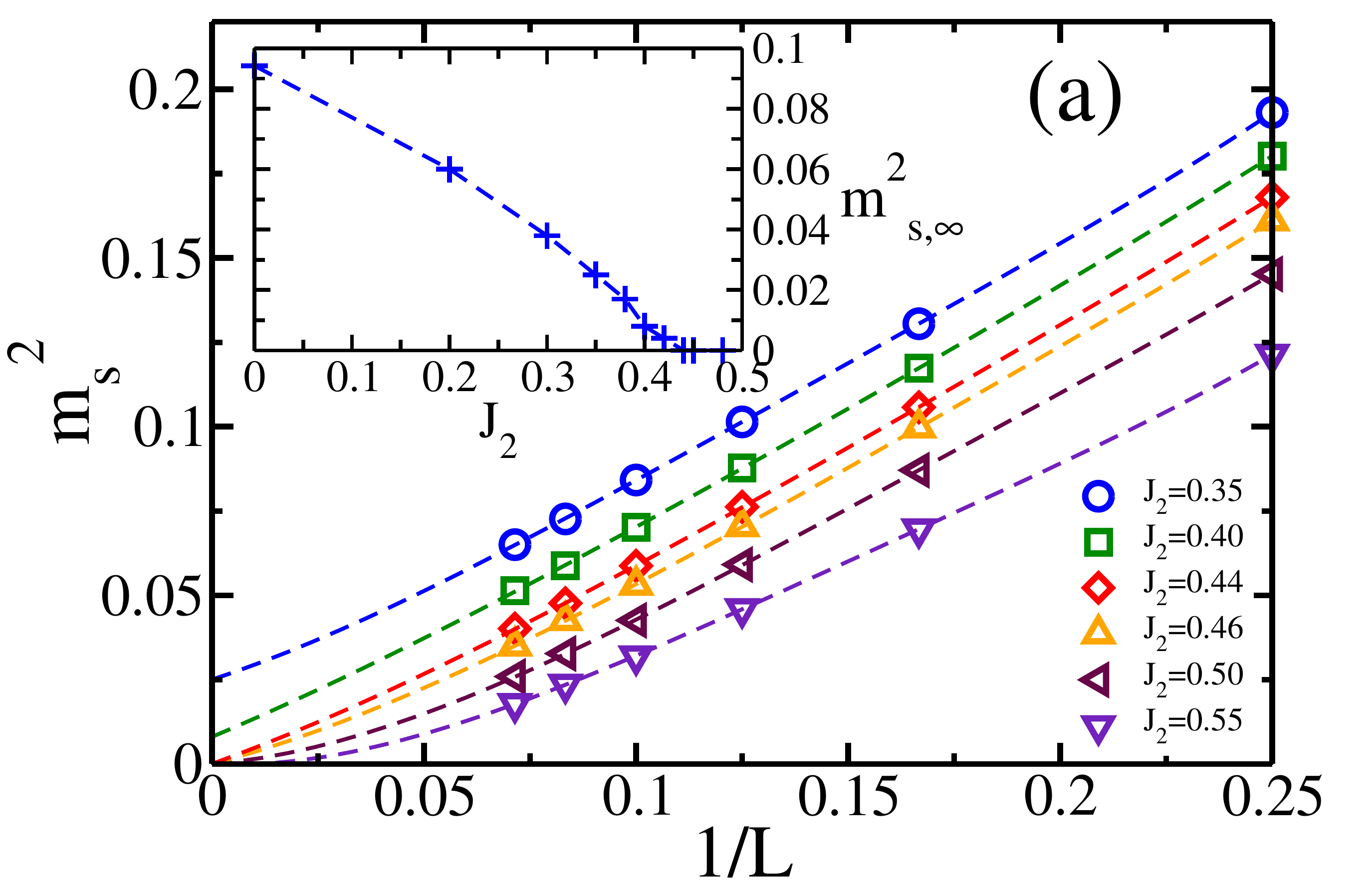}
\includegraphics[width = 1.0\linewidth,clip]{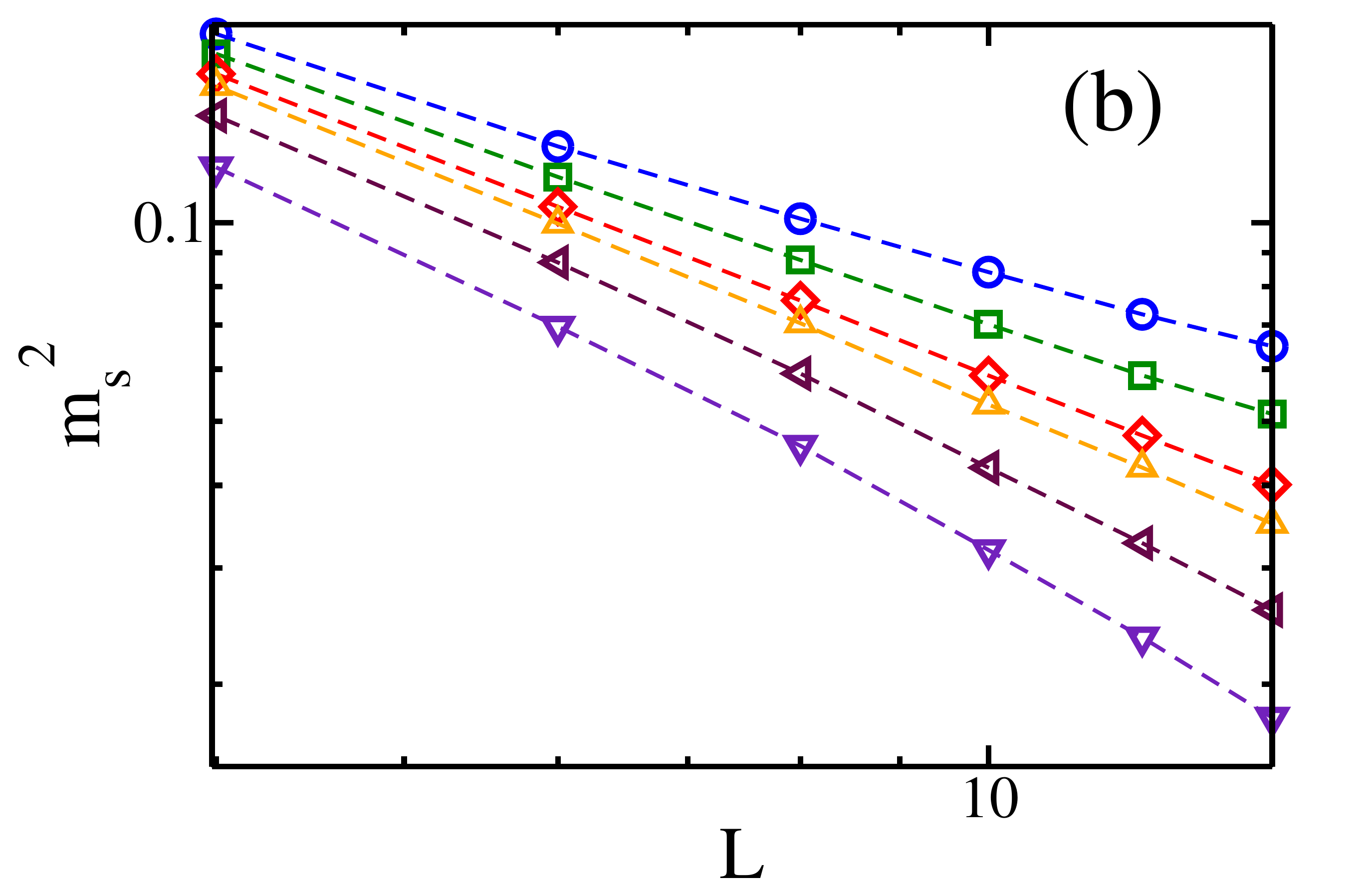}
\caption{(color online) (a) $m_{s}^{2}$ plotted versus $1/L$ for RC$L$-$2L$ cylinder with $L = 4,6,8,10,12,14$; lines are polynomial fits up to fourth order. The inset is $J_2$ dependence of the obtained magnetic order in the 2D limit $m_{s,\infty}^{2}$. (b) Same data as (a) shown as log-log plot of $m_{s}^{2}$ versus width $L$.}\label{spin_order}
\end{figure}

\textit{N\'{e}el order.---}The N\'{e}el order parameter $m_{s}^{2}$ is defined as $m_{s}^{2}=\frac{1}{N^2}\sum_{i,j}\langle S_{i}\cdot S_{j}\rangle e^{i\vec{q}\cdot(\vec{r}_i-\vec{r}_j)}$ ($N$ is the total site number), with $\vec{q} = (\pi,\pi)$.
We calculate $m_{s}^{2}$ from the spin correlations of the $L \times L$ sites in the middle of the RC$L$-$2L$ cylinder, which efficiently reduces boundary effects\cite{PRB_86_024424,PRL_99_127004}.
In Fig.~\ref{spin_order}(a), we show $m_{s}^{2}$ for different systems with $L = 4 - 14$\cite{Neel}.
We show the obtained two-dimensional (2D) limit $m_{s, \infty}^{2}$ in the inset of Fig.~\ref{spin_order}(a).
Such an analysis suggests that the N\'{e}el order vanishes for $J_2 > 0.44$.

The estimated $J_2$ of spin order vanishing is different from the point $J_2 = 0.5$ where the PVB order
develops as found below.  One possibility is an intermediate SL phase\cite{PRL_111_037202,PRB_88_060402}.  Another possibility is that the system is near critical for $0.44 < J_2 < 0.5$.  In this case, to get some idea about the criticality, Fig.~\ref{spin_order}(b) shows the log-log plot of $m_{s}^{2}(L)$.  $m_{s}^{2}$ approaches finite value in the N\'{e}el phase %, and we see this developing
as seen for $J_2 = 0.35$ and $0.4$. 
On the other hand, we expect $m_{s}^{2}(L) \sim L^{-(1 + \eta)}$ at a critical point and $m_{s}^{2}(L) \sim L^{-2}$ in the nonmagnetic phase. The accelerated decay of $m_s^{2}(L)$ at $J_2 = 0.55$ is consistent with vanishing N\'{e}el order: from the two largest sizes we estimate $m_{s}^{2}(L) \sim L^{-1.82}$, which is quite close to $m_{s}^{2}(L) \sim L^{-2}$.
In the near-critical region, we fit the $J_2 = 0.44$ data to $L^{-(1+0.15)}$ and the $J_2 = 0.5$ data ($L > 8$) to $L^{-(1 + 0.44)}$. This range of $\eta$ is compatible with the findings in the $J$-$Q$ models on the square ($\eta \simeq 0.26 - 0.35$)\cite{PRL_98_227202, PRL_100_017203, JSM_2008_02009, PRB_80_180414, PRB_80_212406, PRL_104_177201,PRL_111_137202} and honeycomb ($\eta \simeq 0.3$)\cite{PRL_111_087203} lattices, which show continuous N\'{e}el-to-VBC transition argued to be in the deconfined criticality class, so our model is compatible with this scenario as well.

\begin{figure}[tbp]
\includegraphics[width = 1.0\linewidth,clip]{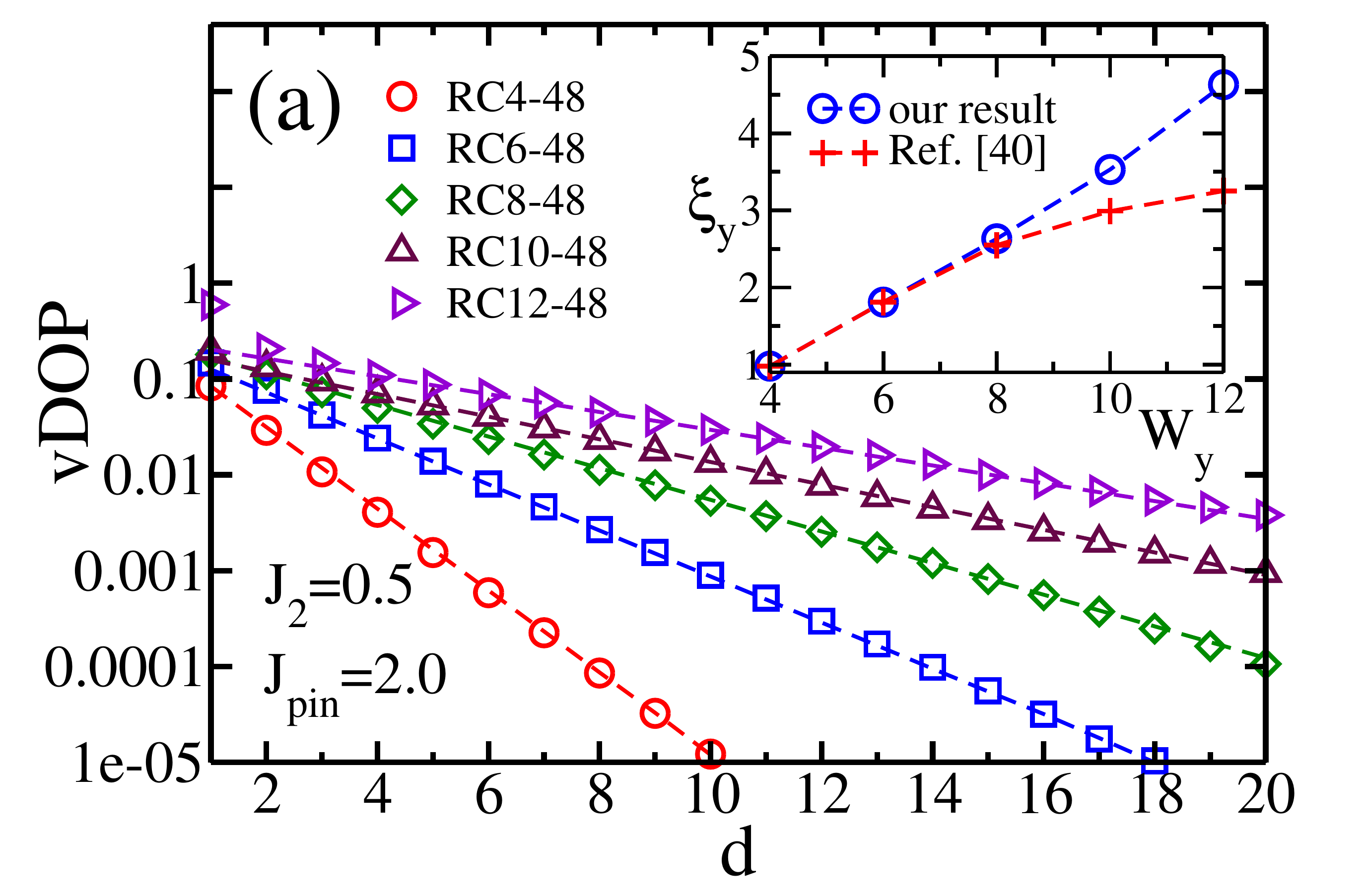}
\includegraphics[width = 1.0\linewidth,clip]{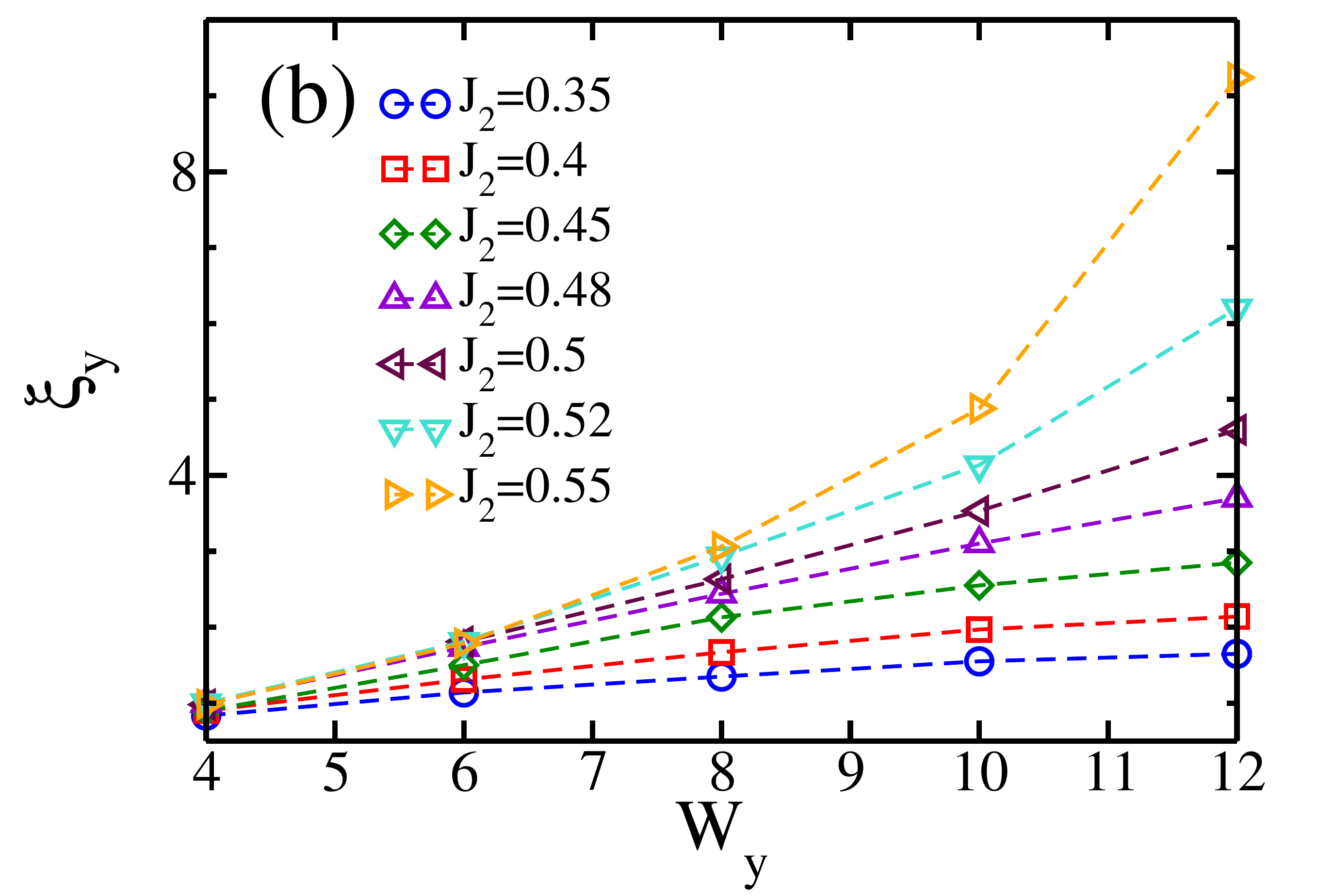}
\includegraphics[width = 1.0\linewidth,clip]{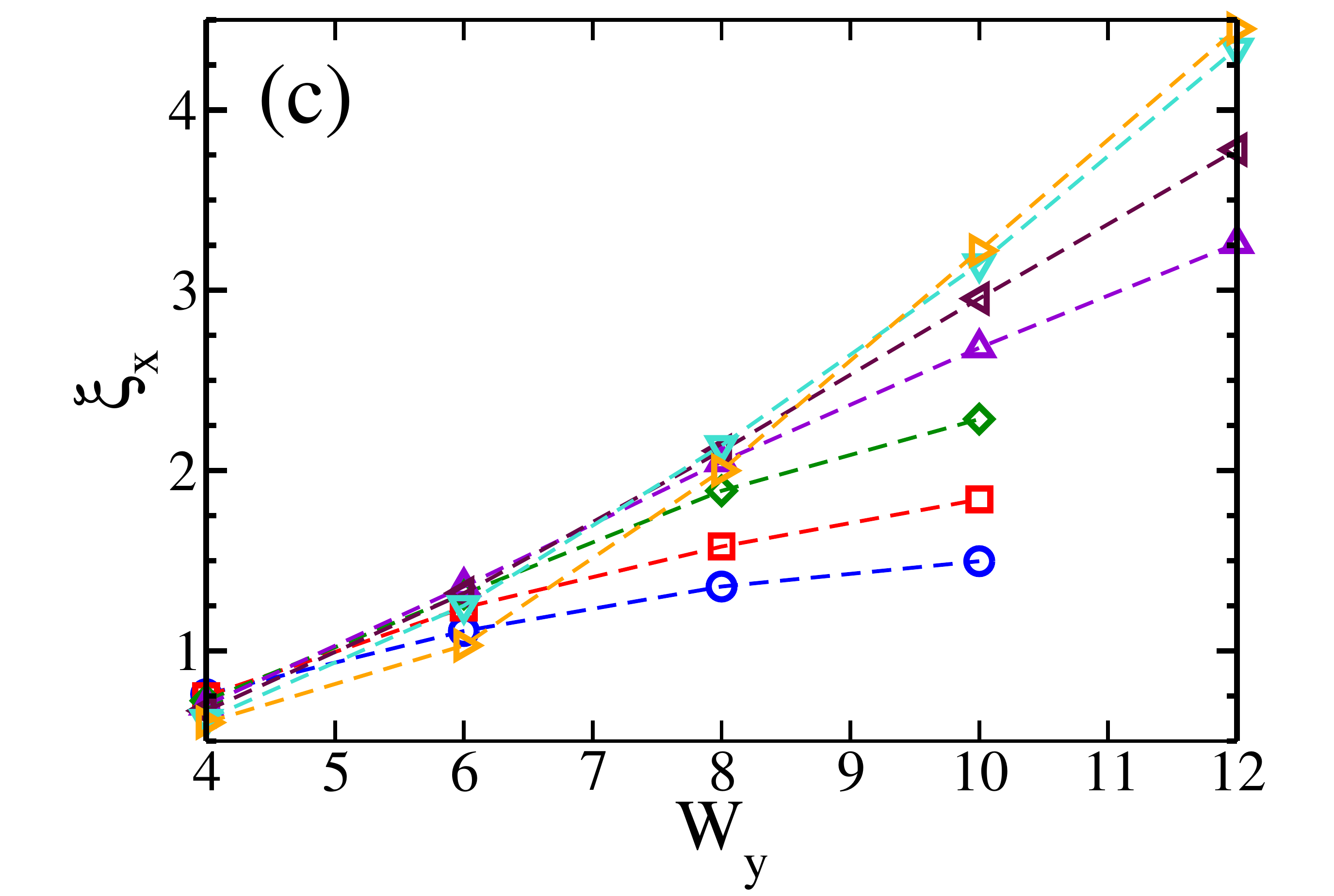}
\caption{(color online) (a) Log-linear plot of vDOP for $J_2 = 0.5$ and $J_{\rm pin} = 2.0$ on the RC cylinder. The inset is the comparison of width dependence of the vertical dimer decay length $\xi_{\rm y}$ with Ref.~\cite{PRB_86_024424}. (b), (c) $\xi_{\rm y}$ and $\xi_{\rm x}$ versus $W_y$ on RC cylinders with $J_{\rm pin} = 2.0$ for a range of $J_2$ shown with the same symbols in both panels.}
\label{dimer_RC}
\end{figure}

\textit{VBC orders.---}We introduce the ``pinning'' bonds $J_{\rm pin} \neq J_1$ on boundaries to induce a vertical dimer pattern, and measure the decay length of the dimer order parameter (DOP) texture from the edge to the middle of the cylinder\cite{PRB_86_024424,PRB_85_134407}.
The vertical DOP (vDOP) is defined as the difference between the strong and weak vertical bond energies.
In Fig.~\ref{dimer_RC}(a), we show a log-linear plot of the vDOP for $J_2 = 0.5$ and $J_{\rm pin} = 2.0$ on long cylinders.
We find that, although the amplitude of the vDOP texture changes with $J_{\rm pin}$, the decay length $\xi_{\rm y}$ is independent of $J_{\rm pin}$ (see Supplemental Material\cite{Sup}).
In the inset of Fig.~\ref{dimer_RC}(a) we compare our $\xi_{\rm y}$ with those in Ref.~\cite{PRB_86_024424}.  We observe consistency for $W_y \leq 8$, but disagreement for $W_y \geq 10$\cite{discrepancy}.
The disagreement might originate from less good convergence in Ref.~\cite{PRB_86_024424}.
Our results are fully converged by keeping 16000 (24000) states for $L_y = 10$ ($12$) systems.
In Fig.~\ref{dimer_RC}(b), we show the width dependence of $\xi_{\rm y}$ for various $J_2$.
$\xi_{\rm y}$ grows slowly and saturates on wide cylinders for $J_2 < 0.5$, demonstrating the vanishing VBC order.
For $J_2 > 0.5$, $\xi_{\rm y}$ grows faster than linear, suggesting nonzero vDOP in the 2D limit.
In addition to the vertical dimer, the system also has the horizontal bond dimer with an exponentially decaying horizontal DOP (hDOP).
In Fig.~\ref{dimer_RC}(c), we show that the hDOP decay length $\xi_{\rm x}$ also grows strongly for $J_2 > 0.5$.
The coexisting nonzero horizontal and vertical dimer orders suggest a PVB state.

\begin{figure}[tbp]
\includegraphics[width = 1.0\linewidth,clip]{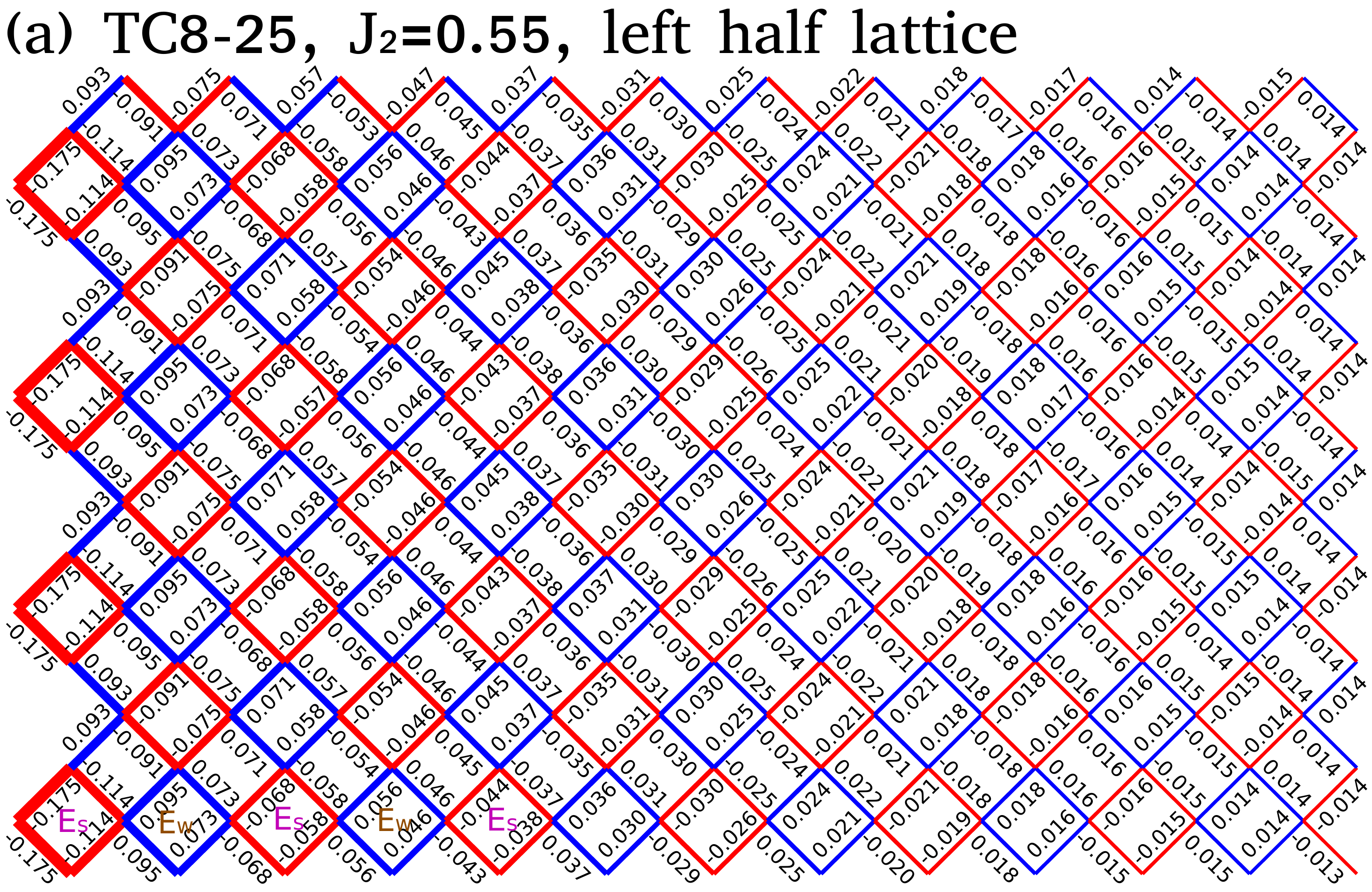}
\includegraphics[width = 1.0\linewidth,clip]{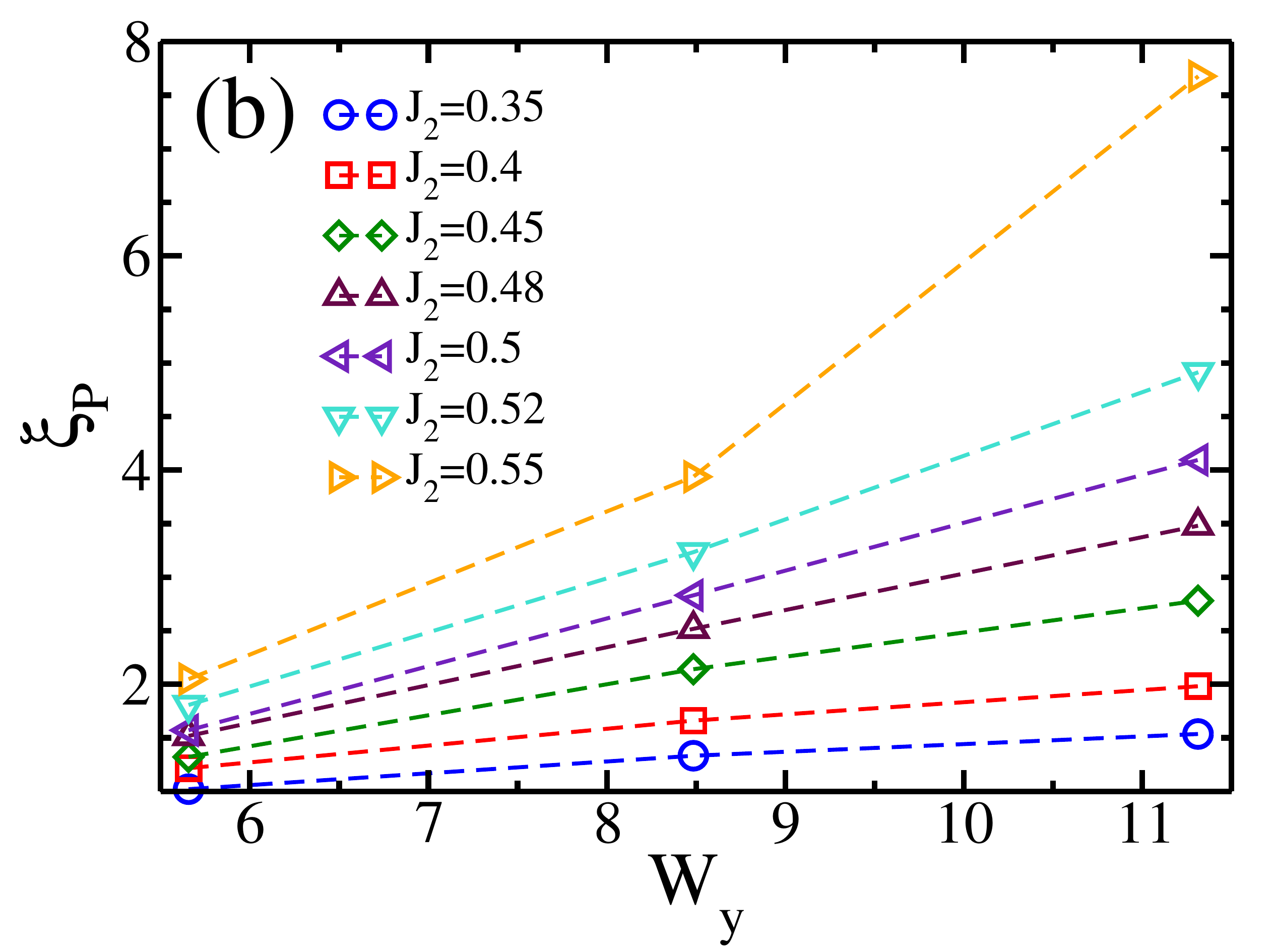}
\caption{(color online) (a) The NN $J_1$ bond energy for $J_2 = 0.55$ on the left half of the tilted TC8-25 cylinder, where we have subtracted a constant $-0.2948$ from all the bond energies.  Note that the TC cylinders have the square lattice rotated by 45 deg compared to Fig.~\ref{phase_diagram}.
We trim every other site on both boundaries to make lattice select unique PVB order. $E_{s}$ ($E_{w}$) denotes the sum of four NN bond energies of the red (blue) plaquette with negative (positive) numbers. (b) Dependence of the pDOP decay length $\xi_{\rm P}$ on the cylinder width $W_y$.
}\label{dimer_TC}
\end{figure}

We also study the dimer structure factors $S_{\rm VBC}$ and $S_{\rm col}$ defined in Ref.~\cite{PRB_74_144422}; the former detects both the PVB and CVB orders while the latter is nonzero only for the CVB order.
We take RC$L$-$2L$ cylinders without pinning and calculate the structure factors using the dimer correlations of the middle $L \times L$ sites.
The picture of the dimer correlations is consistent with the $s$-wave plaquette state\cite{PRB_74_144422}. The finite-size extrapolations show that while $S_{\rm VBC}/N$ possibly approaches finite values for $J_2 > 0.5$, $S_{\rm col}/N$ clearly approaches zero, which definitely excludes the CVB order.

To explicitly demonstrate PVB order, we study the TC obtained by cutting the cylinder along the $J_2$ direction and trimming every other site on the boundary as shown in Fig.~\ref{dimer_TC}(a).
We label it as TC$L_y$-$L_x$, where $L_y$ and $L_x$ denote the number of square plaquettes along the $y$ and $x$ directions; the width of the cylinder is $W_y = \sqrt{2} L_y$.
The trimmed edges induce strong PVB order on the boundaries.
We denote the sum of the four NN bond energies of a ``strong'' red (``weak'' blue) plaquette as $E_s$ ($E_w$),
and define the plaquette DOP (pDOP) as the difference between $E_s$ and $E_w$, which is found to decay exponentially with a decay length $\xi_{\rm P}$. In Fig.~\ref{dimer_TC}(b), we find strong growth of $\xi_{\rm P}$ with $W_y$ for $J_2 > 0.5$, consistent with a PVB state.
By studying the log-log plot of VBC order parameter versus system size (see Supplemental Material\cite{Sup}), we estimate the anomalous exponent of dimer correlations $\eta_{\rm VBC} \simeq 0.4$ at $J_2 = 0.5$, which is not far from estimates in the deconfined criticality scenario in the $J$-$Q$ models
on square ($\eta_{\rm VBC} \simeq 0.25$)\cite{PRL_98_227202, PRL_100_017203, JSM_2008_02009, PRB_80_180414, PRB_80_212406, PRL_104_177201, PRL_111_137202} and honeycomb ($\eta_{\rm VBC} \simeq 0.28$)\cite{PRL_111_087203} lattices.

\begin{figure}[tbp]
\includegraphics[width = 1.0\linewidth,clip]{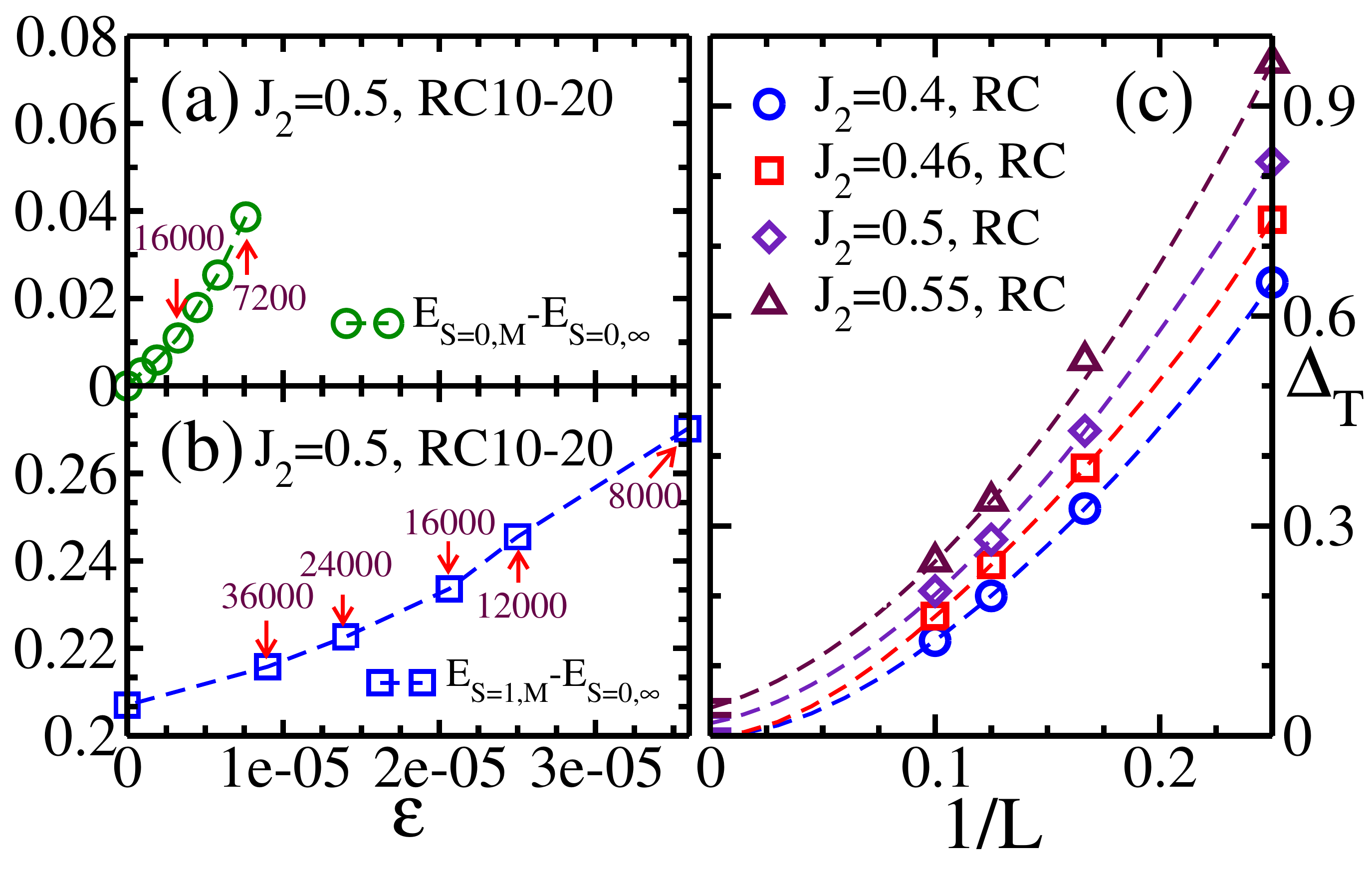}
\caption{(color online) (a), (b) Ground-state energies for RC10-20 cylinder at $J_2 = 0.5$ in the $S=0$ ($E_{S=0,M}$) and $S=1$ ($E_{S=1,M}$) sectors versus the DMRG truncation error $\varepsilon$.  All the energies have subtracted the ground-state energy $E_{S=0,\infty} = -99.022(1)$. $M$ is the number of kept $U(1)$-equivalent DMRG states and is indicated next to the symbols.
(c) Finite-size extrapolations of spin gap $\Delta_{\rm T}$ on RC$L$-$2L$ cylinders ($L = 4,6,8,10$). 
For $J_2 < 0.5$, the data are fitted using the formula $\Delta_{\rm T}(L) = \Delta_{\rm T}(\infty) + \alpha/L^2 + \beta/L^3 +\gamma/L^4$, while for $J_2 \geq 0.5$, we fit the data using $\Delta_{\rm T}(L) = \Delta_{\rm T}(\infty) + a/L + b/L^2 + c/L^4$.
We estimate $\Delta_{\rm T}(\infty) = 0.018\pm 0.01$ and $0.04\pm 0.01$ for $J_2 = 0.5$ and $0.55$, respectively. } \label{spin_gap}
\end{figure}

\textit{Spin gap and ground state energy.---}We calculate the spin gap $\Delta_{\rm T}$ on the RC$L$-$2L$ cylinders up to $L = 10$ following the method from Ref.~\cite{Science_332_1173}: We sweep the ground state first, and then target the $S=1$ sector sweeping the middle $L \times L$ sites to avoid edge excitations.
In Figs.~\ref{spin_gap}(a) and \ref{spin_gap}(b), we show energies versus the DMRG truncation error for the RC10-20 cylinder at $J_2 = 0.5$ in the $S=0$ and $S=1$ sectors. In both plots we have subtracted the ground-state energy $-99.022(1)$ obtained by the extrapolation in Fig.~\ref{spin_gap}(a).
We find that we need about twice as many states to achieve
the same energy error in the $S=1$ sector as in the $S=0$ sector.
The difficulty to reach the convergent energy in the $S=1$ sector may explain the overestimate of the spin gap in the earlier work \cite{PRB_86_024424}:  We find $\Delta_{\rm T} \simeq 0.207$ while Ref.~\cite{PRB_86_024424} estimates $\Delta_{\rm T} \simeq 0.248$.
We obtain accurate spin gaps by keeping up to $36000$ states at $L_y = 10$, which sets the limit of our simulations.

Figure~\ref{spin_gap}(c)  shows the finite-size extrapolations of $\Delta_{\rm T}$.
In our fits, we find $\Delta_{\rm T}$ extrapolating vanishing for $J_2 \leq 0.48$, consistent with the N\'{e}el order for $J_2 \leq 0.44$. For $J_2 = 0.5$ and $0.55$, $\Delta_{\rm T}(L\to \infty)$ is fitted to $0.018\pm 0.01$ and $0.04 \pm 0.01$, respectively; this is compatible with a VBC phase.

\textit{Summary and discussion.---}We have studied the ground state of spin-$\frac{1}{2}$ $J_1$-$J_2$ SHM by accurate $SU(2)$ DMRG simulations.  We find a N\'{e}el order persisting up to $J_2 = 0.44$.
Contrary to the previous proposals of gapped $Z_2$ SL from DMRG\cite{PRB_86_024424} or gapless $Z_2$ SL from VMC calculations\cite{PRB_88_060402}, we establish an $s$-wave PVB state for $J_2 > 0.5$ by observing
rapidly growing characteristic lengths of both the vertical and horizontal dimer orders on different cylinders.
Between the N\'{e}el and PVB phases, we find a near-critical region that could be compatible with the deconfined criticality scenario.
However, since the system in this region has large correlation length scales that can be comparable to or even larger than the system widths we can approach, we cannot exclude a possible gapless SL region proposed in variational studies\cite{PRL_111_037202,PRB_88_060402}.
We hope that future studies on larger system size, either pushing DMRG further or using new techniques such as tensor network will be able to resolve between these scenarios more clearly.

We would like to particularly thank H.-C. Jiang and L. Balents for extensive discussions.
We also acknowledge stimulating discussions with K.~S.~D.~Beach, Z.-C.~Gu, W.-J.~Hu, L.~Wang, S.~White, Z.-Y.~Zhu, and A. Sandvik.
This research is supported by the National Science Foundation through grants DMR-1205734 (S.S.G.), DMR-0906816 (D.N.S.), DMR-1206096 (O.I.M.), DMR-1101912 (M.P.A.F.),
the U.S. Department of Energy, Office of Basic Energy Sciences under grant No. DE-FG02-06ER46305 (W.Z),
and by the Caltech Institute of Quantum Information and Matter, an NSF Physics Frontiers Center with support of the Gordon and Betty Moore Foundation (O.I.M. and M.P.A.F.).

\newpage
\begin{appendices}

\begin{center}
\textbf{Supplementary Information}
\end{center}

% % % % % % % % % % % % % % %Ground-state energy
\begin{figure}[tbp]
\includegraphics[width = 1.0\linewidth,clip]{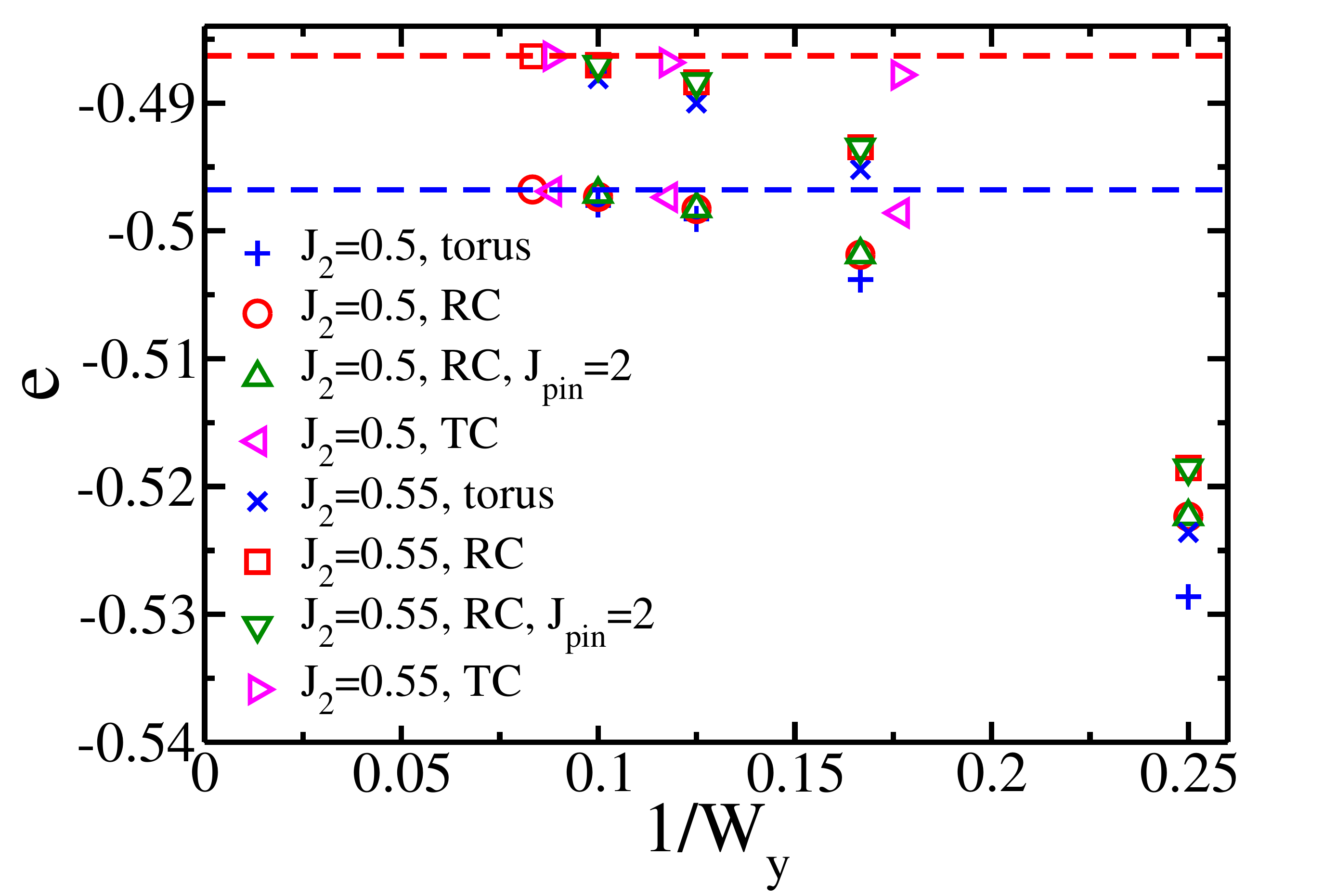}
\caption{(color online) DMRG ground-state energy per site for $J_2 = 0.5$ and $0.55$ on torus, on RC cylinder without pinning or with
vertical dimer boundary pinning $J_{\rm pin} = 2.0$, as well as on TC cylinder. The energies on torus are obtained through extrapolation with DMRG truncation error (see Table~\ref{energy_table}). On cylinder, we get bulk energy by subtracting the energies of two long cylinders with different system lengths. With growing system width, the energies on different samples approach each other, giving the 
estimates of ground-state energy in the 2D limit for $J_2 = 0.5$ and $0.55$ as $e_{\infty} \simeq -0.4968$ and $-0.4863$, respectively.
}\label{energy}
\end{figure}

\textit{DMRG ground-state energies for $J_2 =0.5$ and $0.55$.---}We show our DMRG ground-state energies for $J_2 = 0.5$ and $0.55$ in Fig.~\ref{energy}.
We study $L \times L$ torus systems with $L = 4,6,8,10$.
We keep more than $M=32000$ optimal states for DMRG sweeping, and estimate the energy through extrapolation of finite-$M$ energies via DMRG truncation error (see data in Table \ref{energy_table} below).
For cylinders, we obtain bulk energy by subtracting the energies of two long cylinders with different system lengths to eliminate boundary effects.

As shown in Fig.~\ref{energy}, the energies per site of all samples increase slowly with increasing system width $W_y$ and approach close to each other for $W_y \gtrsim 10$.
The energies on torus are lower than those on cylinder, and the difference decreases with increasing $W_y$.
The bulk energy on RC cylinder is essentially 
independent of the boundary pinning $J_{\rm pin}$.
As the ground-state energy appears close to convergence for $W_y \geq 8$, we take a simple straight line fitting of the large-size results to give estimates of the energy in the 2D limit $e_{\infty}$ as shown by the dashed lines in Fig.~\ref{energy}.
We find $e_{\infty} \simeq -0.4968$ and $-0.4863$ for $J_2 = 0.5$ and $0.55$, respectively.

%%%%%%%%%%%%%%%%%%%%%%%%%%%%%%%%%%%%%%%%%%%%%%%%%%%%%%%%%%%%%%%%%%%%%%%%

\begin{figure}[tbp]
\includegraphics[width = 1.0\linewidth,clip]{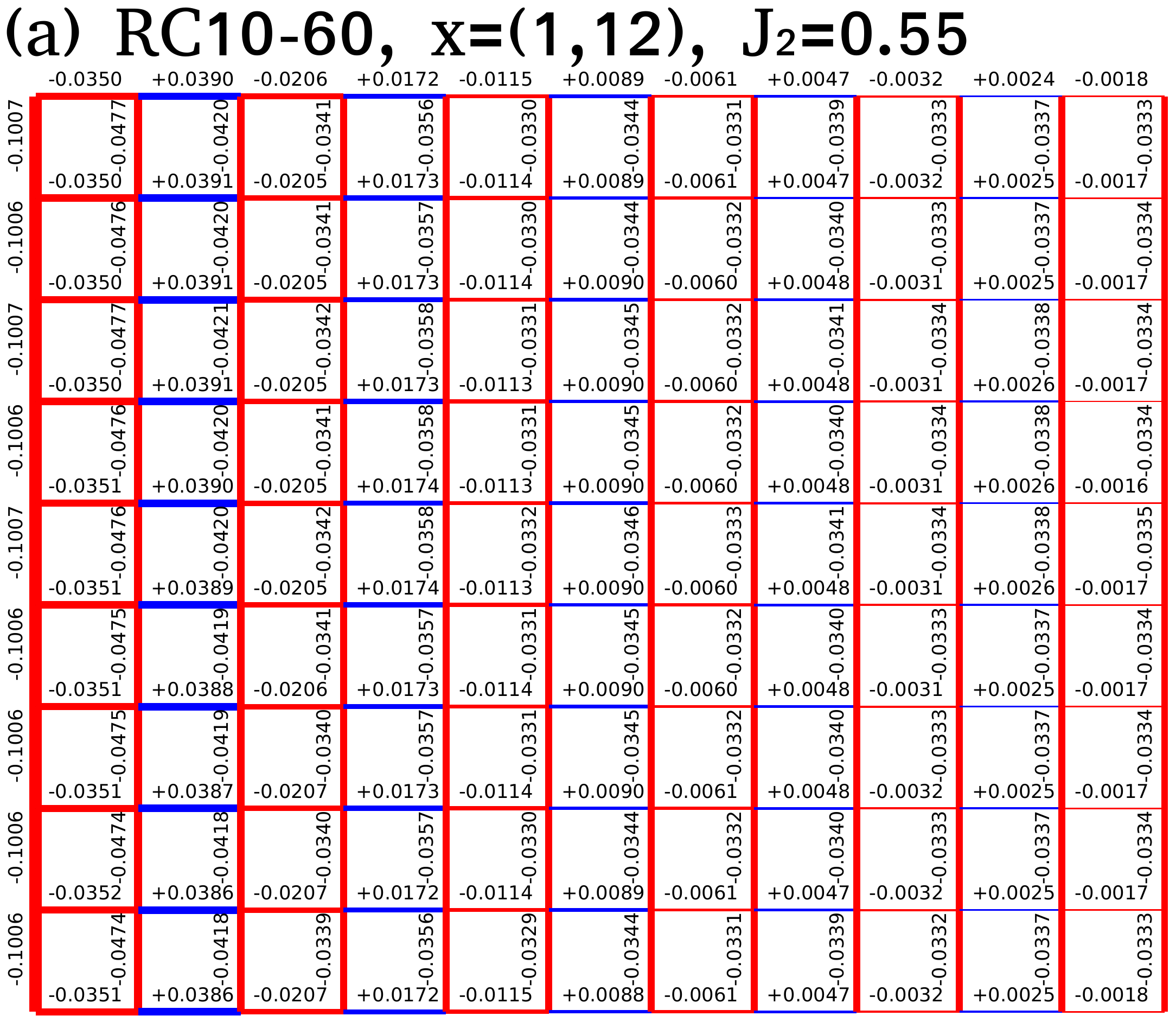}
\includegraphics[width = 1.0\linewidth,clip]{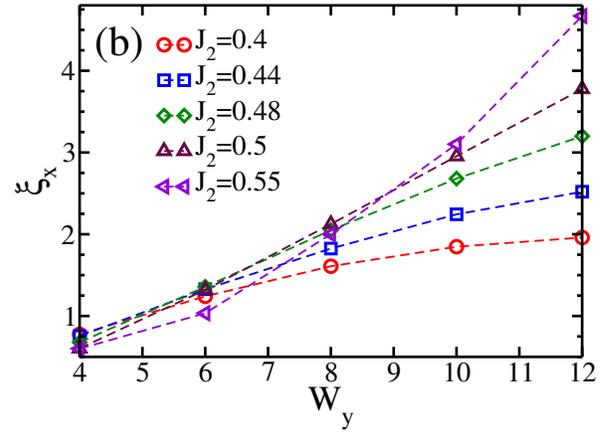}
\caption{(color online) (a) Subtracted NN bond energies for $J_2 = 0.55$ on RC10-60 cylinder without pinning; the subtracted number $-0.2763$ is the average horizontal bond energy in the bulk of the lattice.  Here we show the left $12$ columns.  The alternation of red (negative number) and blue (positive number) bonds indicates horizontal dimer texture. (b) hDOP decay length $\xi_{\rm x}$ versus system width on RC cylinder without pinning. The extracted decay lengths $\xi_{\rm x}$ are similar to those in Fig. 3(c) of the main text obtained on RC cylinder with the vertical dimer pinning.}\label{hDOP_nopin}
\end{figure}

\textit{Horizontal dimer order on RC cylinder without pinning.---}On RC cylinder without pinning, the open edges break the lattice translation symmetry only in the $x$ direction.
The horizontal nearest-neighbor (NN) bond energies have the ``strong-weak" dimer pattern as shown in Fig.~\ref{hDOP_nopin}(a). We define the hDOP as the difference of the adjacent horizontal NN bond energies, which decays exponentially with a decay length $\xi_{\rm x}$. In Fig.~\ref{hDOP_nopin}(b), we show the hDOP decay length $\xi_{\rm x}$ versus system width. For $J_2 < 0.5$, $\xi_{\rm x}$ grows more slowly than linearly and approaches saturation on large size, which is consistent with vanishing dimer order. However, for $J_2 > 0.5$, $\xi_{\rm x}$ grows fast, suggesting nonzero bulk hDOP on wider cylinders. This horizontal dimer order supports our claim of the VBC state for $J_2 > 0.5$. We also find that the $\xi_{\rm x}$ obtained here are almost the same as those in Fig. 3(c) of the main text where the cylinder systems have the vertical bond pinning.

\begin{figure}[tbp]
\includegraphics[width = 1.0\linewidth,clip]{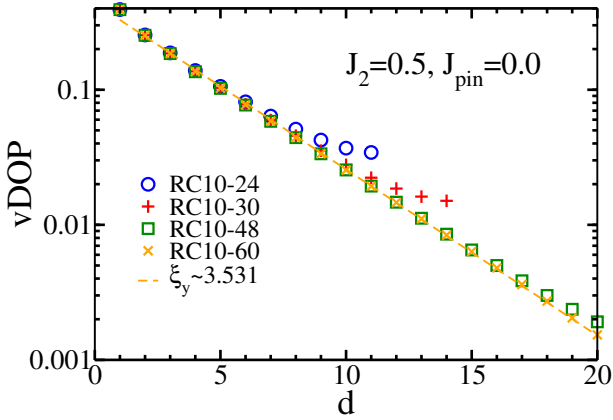}
\caption{(color online) Log-linear plot of the vDOP on RC10 cylinders with different system lengths at $J_2=0.5$ and $J_{\rm pin}=0.0$. The exponential fitting gives decay length $\xi_{\rm y}=3.531$.}\label{length_Lx}
\end{figure}

\begin{figure}[tbp]
\includegraphics[width = 0.495\linewidth,clip]{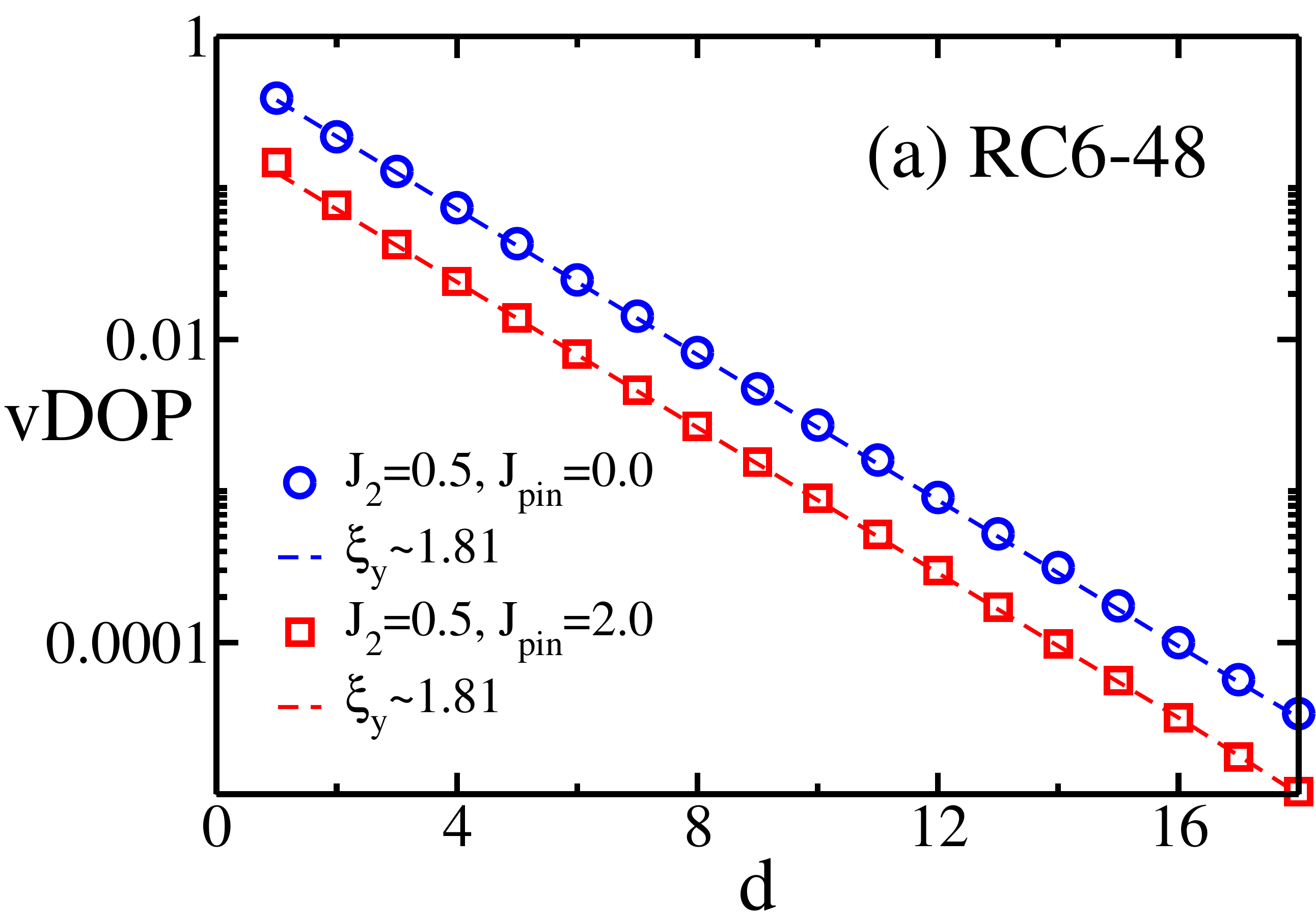}
\includegraphics[width = 0.495\linewidth,clip]{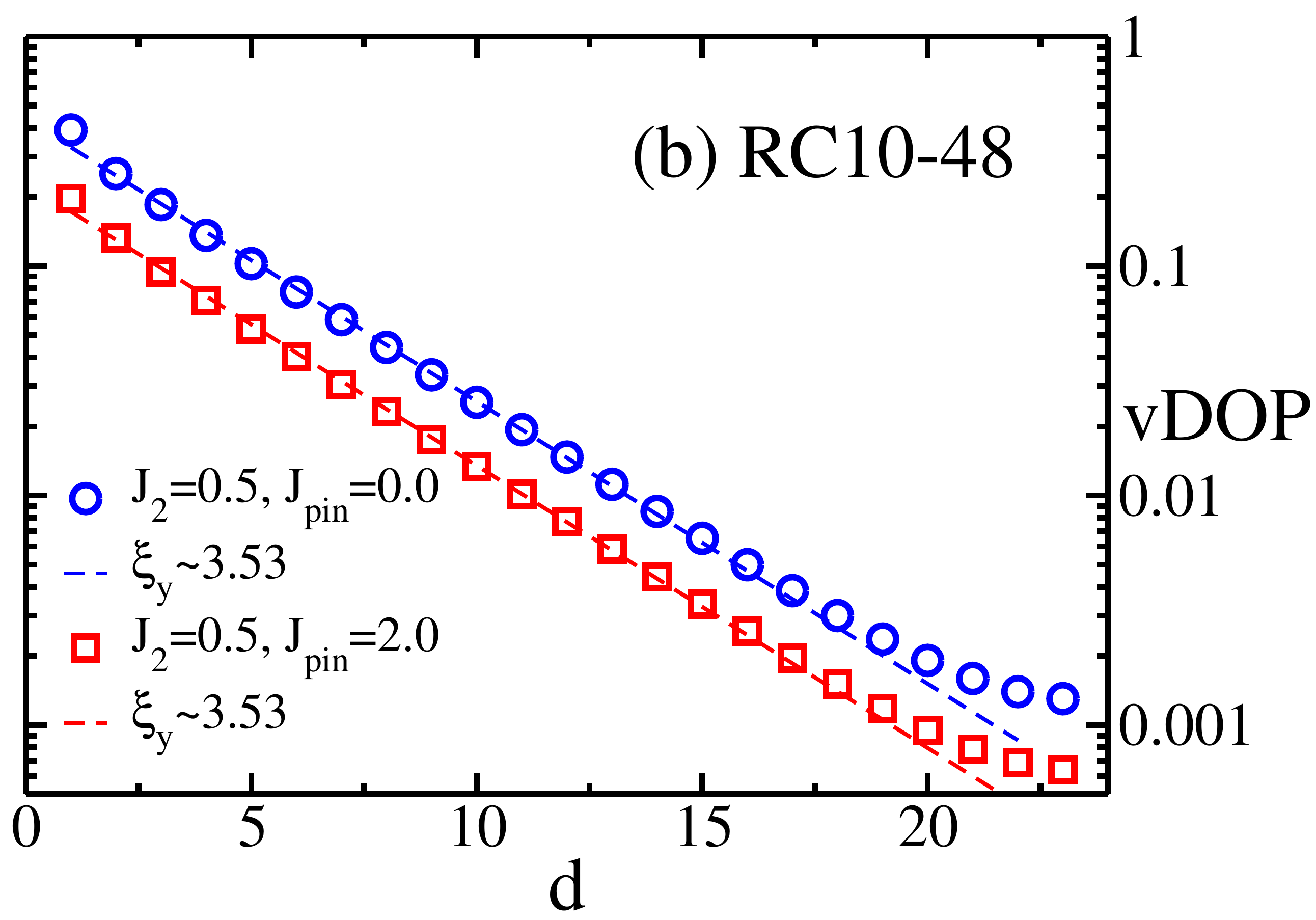}
\includegraphics[width = 0.495\linewidth,clip]{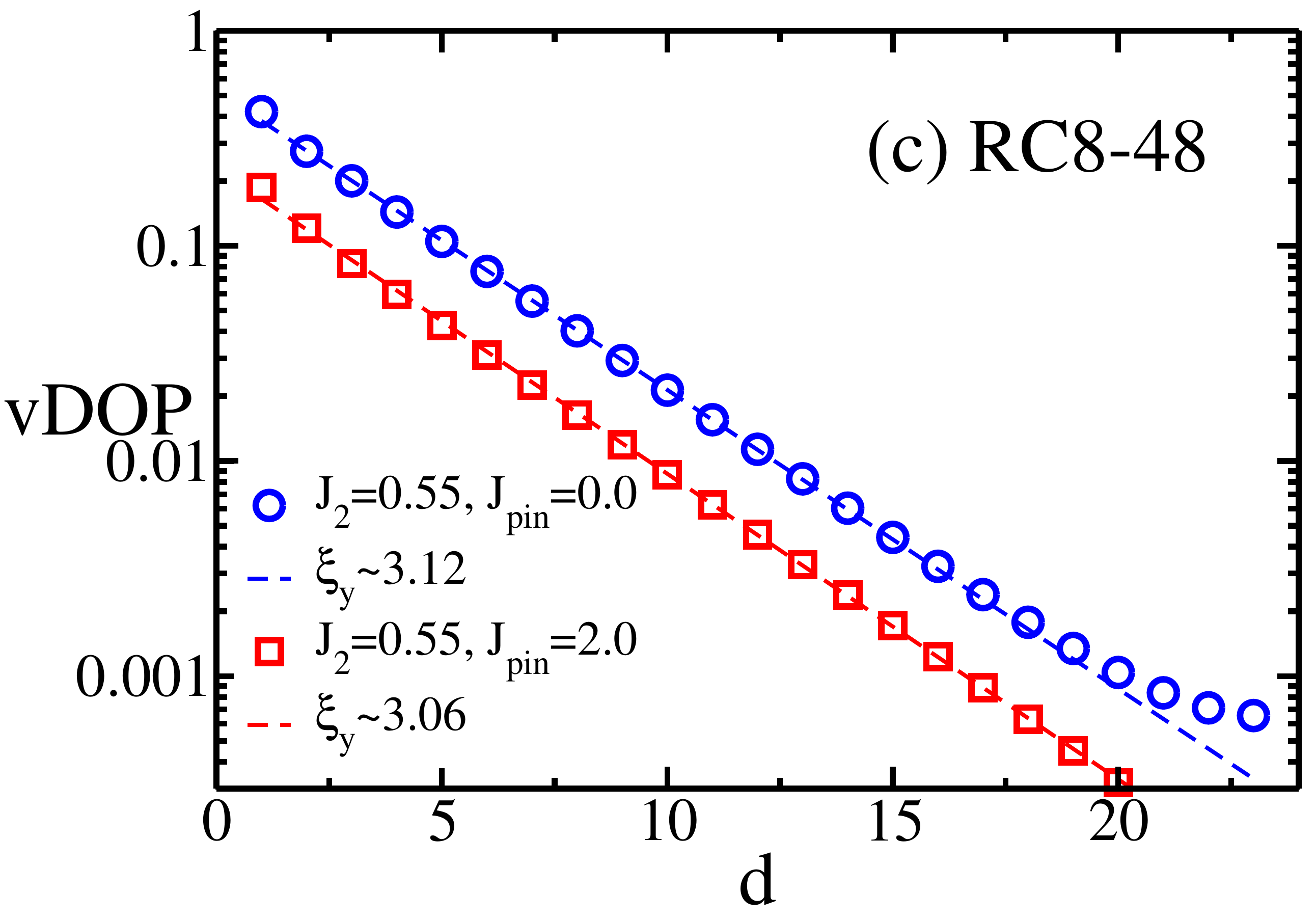}
\includegraphics[width = 0.495\linewidth,clip]{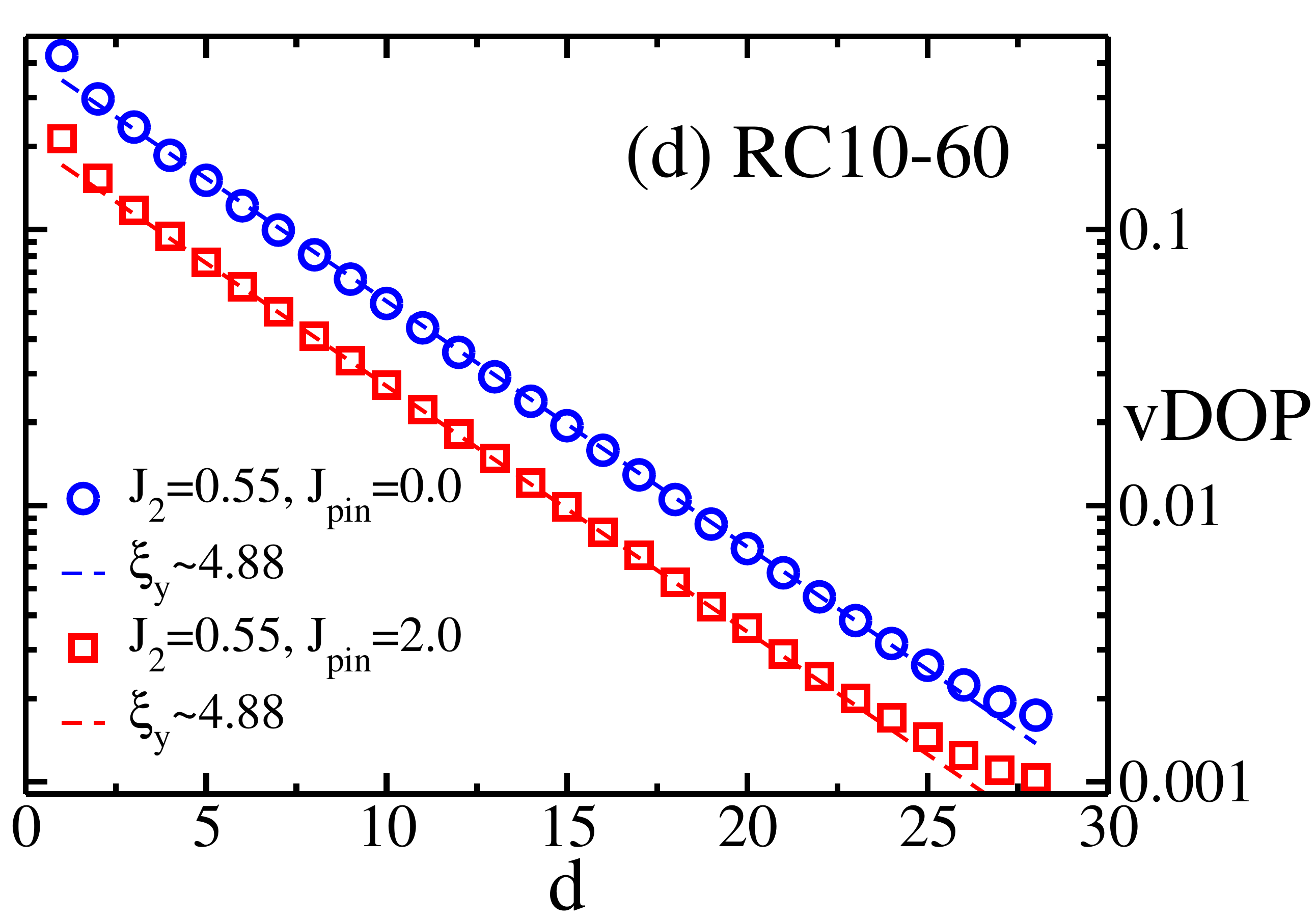}
\caption{(color online) Comparisons of the vDOP textures on RC cylinder with different boundary pinnings. We have studied several different $J_{\rm pin}$ and found that although the vDOP varies with $J_{\rm pin}$, the decay length $\xi_{\rm y}$ is almost independent of the pinning strength. Here we show the results with $J_{\rm pin} = 0.0$ and $2.0$.}\label{pinning}
\end{figure}

\textit{Pinning independence of the vDOP decay length $\xi_{\rm y}$.---} In the main text, we introduced modified vertical bonds $J_{\rm pin}$ on boundaries to break the lattice translational symmetry in the $y$ direction, allowing us to study the vDOP and the width dependence of the vDOP decay length. A direct question is whether the pinning strength
affects these quantities. We have compared the vDOP and its decay length for several different pinning strengths, from weak pinning $J_{\rm pin} = 1.01,1.1,1.2$ to strong pinning $J_{\rm pin} = 2.0$ and $J_{\rm pin} = 0.0$. First of all, in Fig.~\ref{length_Lx} we show that our results are obtained on quite long cylinders, thus minimizing the influence of finite-size effects on the decay length. Next, in Fig.~\ref{pinning} we show some examples of varying boundary pinning at $J_2 = 0.5, 0.55$; we find that although the amplitude of the vDOP texture varies with $J_{\rm pin}$, the decay length $\xi_{\rm y}$ is almost independent of the pinning strength, indicating that our results with pinning are robust properties of the bulk (infinitely long cylinder) phase.

\begin{figure}[tbp]
\includegraphics[width = 1.0\linewidth,clip]{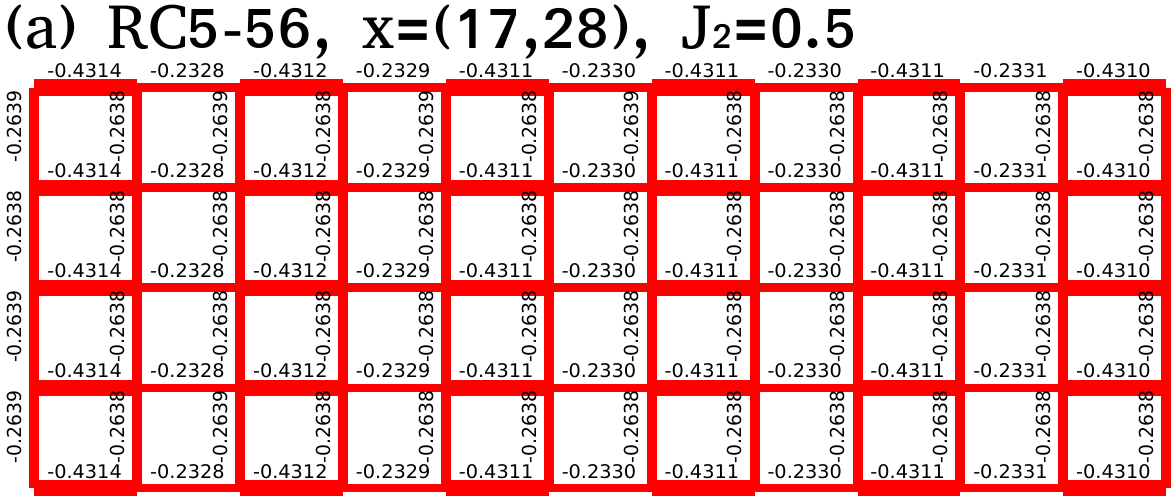}
\includegraphics[width = 1.0\linewidth,clip]{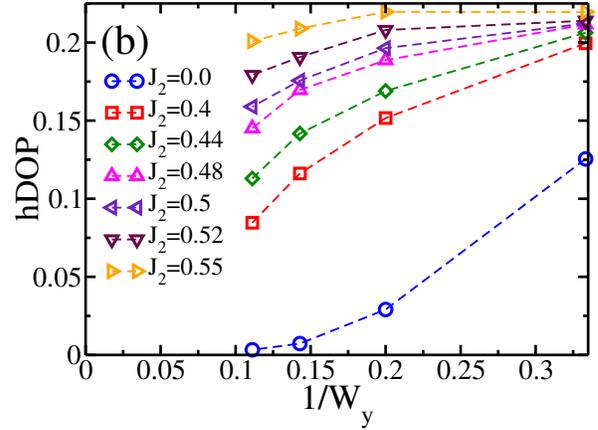}
\caption{(color online) (a) The NN bond energies for RC5-56 cylinder at $J_2 = 0.5$, showing bonds with $x$ from $17$ to $28$.
The system has a spontaneous bulk horizontal dimer order, and the bulk hDOP is defined as the difference of the strong and weak bond energies in the middle of cylinder. (b) Width dependence of the hDOP on the odd-$L_y$ RC cylinders, showing the data for $W_y=3,5,7,9$.}\label{odd}
\end{figure}

\textit{Horizontal dimer order on RC cylinder with odd $L_y$.---}On finite-size odd-$L_y$ RC cylinder, the system spontaneously  develops a nonzero horizontal dimer order in the bulk, which happens both when the 2D phase is VBC or $Z_2$ SL\cite{PRB_86_024424,PRL_108_247206}. For a $Z_2$ SL in the 2D limit, the dimer order would decay exponentially with growing $L_y$. On the other hand, for a VBC state, it should go to a finite value in the 2D limit\cite{PRB_86_024424,PRL_108_247206}. We study the horizontal dimer order on odd-$L_y$ RC cylinder with $L_y$ up to $9$ and $L_x$ up to $100$ to get the results representing $L_x \rightarrow \infty$ cylinders.
We define the absolute difference of the strong and weak horizontal bond energies in the bulk as hDOP, see Fig.~\ref{odd}(a).  We show thus measured hDOP versus $1/W_y$ in Fig.~\ref{odd}(b). For $J_2 < 0.5$ the hDOP decays fast with the cylinder width and appears to extrapolate to zero, while for $J_2 > 0.5$ the hDOP has a slow decay and seems to saturate to a finite value. The nonzero hDOP does not support a $Z_2$ SL, but indicates a VBC state for $J_2 > 0.5$.

\begin{figure}[tbp]
\includegraphics[width = 1.0\linewidth,clip]{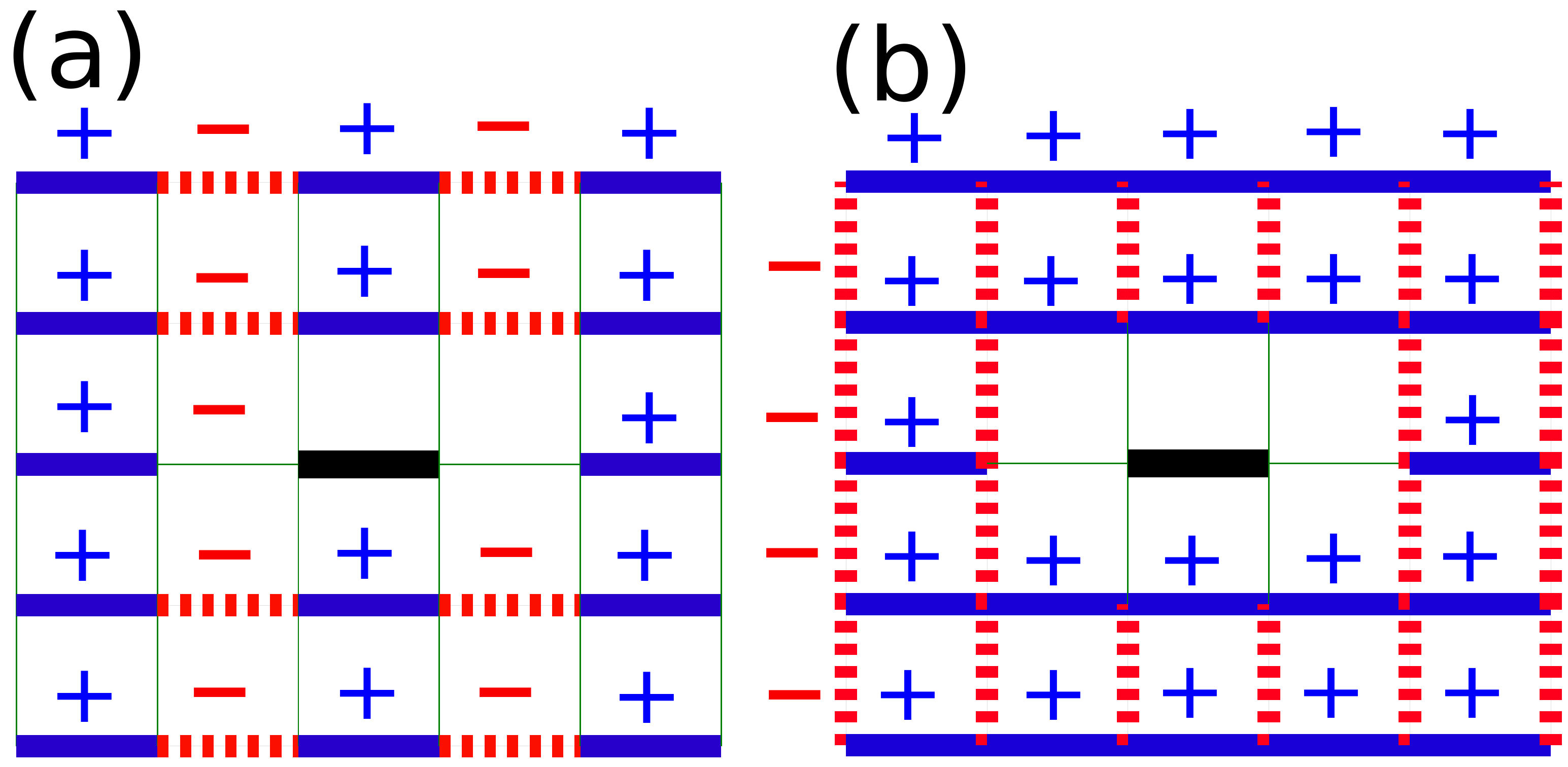}
\caption{(color online) Phase factors $\varepsilon_{\lambda}(k,l)$ for (a) ``vbc" and (b) ``col" dimer structure factors defined in Eq.~(\ref{dimer_stru}). The solid bonds have phase factor $\varepsilon_{\lambda}(k,l) = 1$, while dashed bonds have $\varepsilon_{\lambda}(k,l) = -1$. The $(k,l)$ bonds that are nearest neighbors to the reference bond $(i,j)$ (central black solid bond) are omitted in the calculation of the structure factors.}\label{definition}
\end{figure}

\begin{figure}[tbp]
\includegraphics[width = 1.0\linewidth,clip]{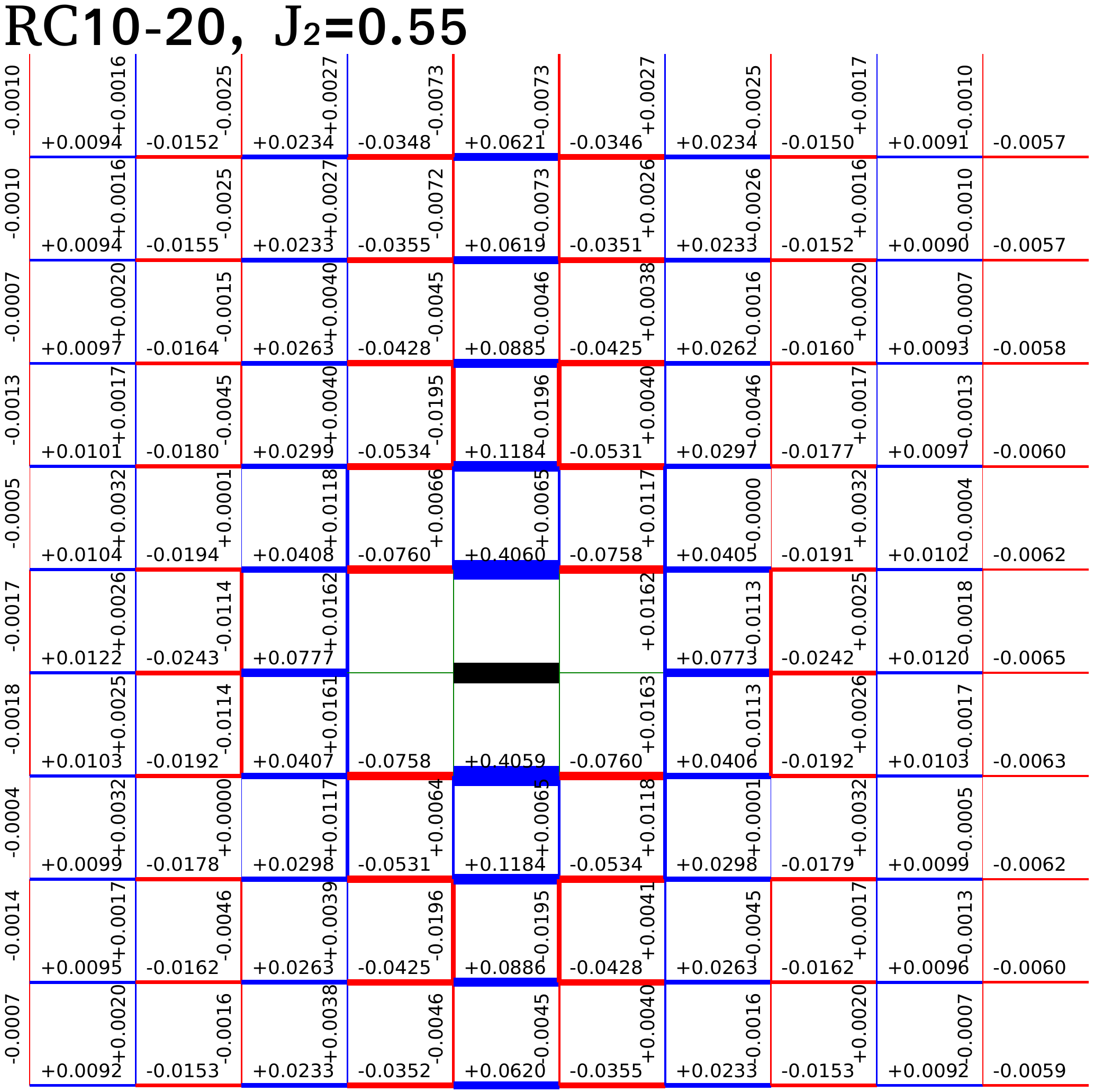}
\caption{(color online) Dimer-dimer correlation function for $J_2 = 0.55$ on RC10-20 cylinder. The black bond in the middle of the cylinder denotes the reference bond $(i,j)$. The blue and red bonds indicate the positive and negative correlations, respectively. Here the middle $10 \times 10$ lattice dimer correlations are shown, which are used to calculate the dimer structure factors.}\label{dimer_pattern}
\end{figure}

\begin{figure}[tbp]
\includegraphics[width = 1.0\linewidth,clip]{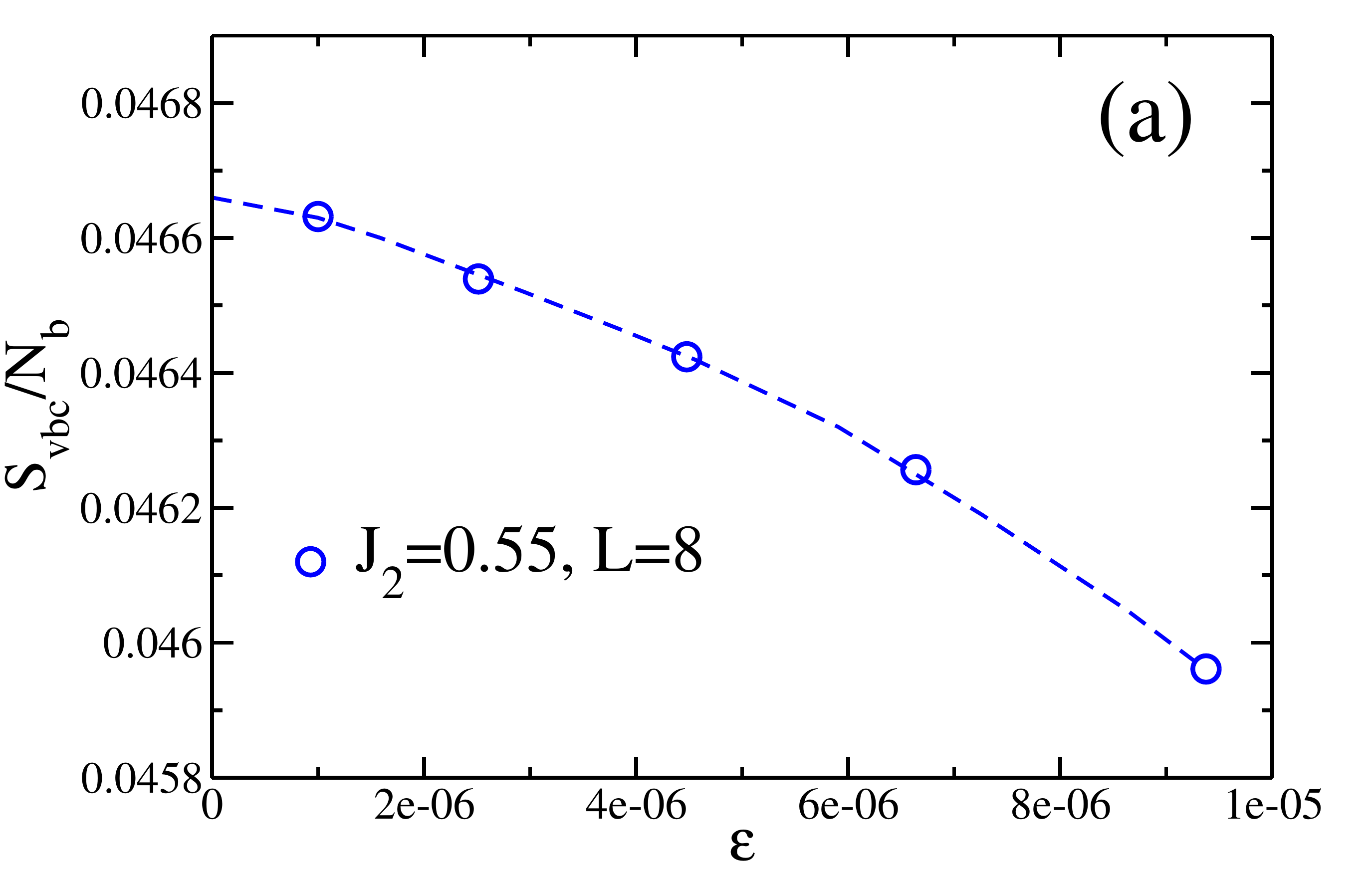}
\includegraphics[width = 1.0\linewidth,clip]{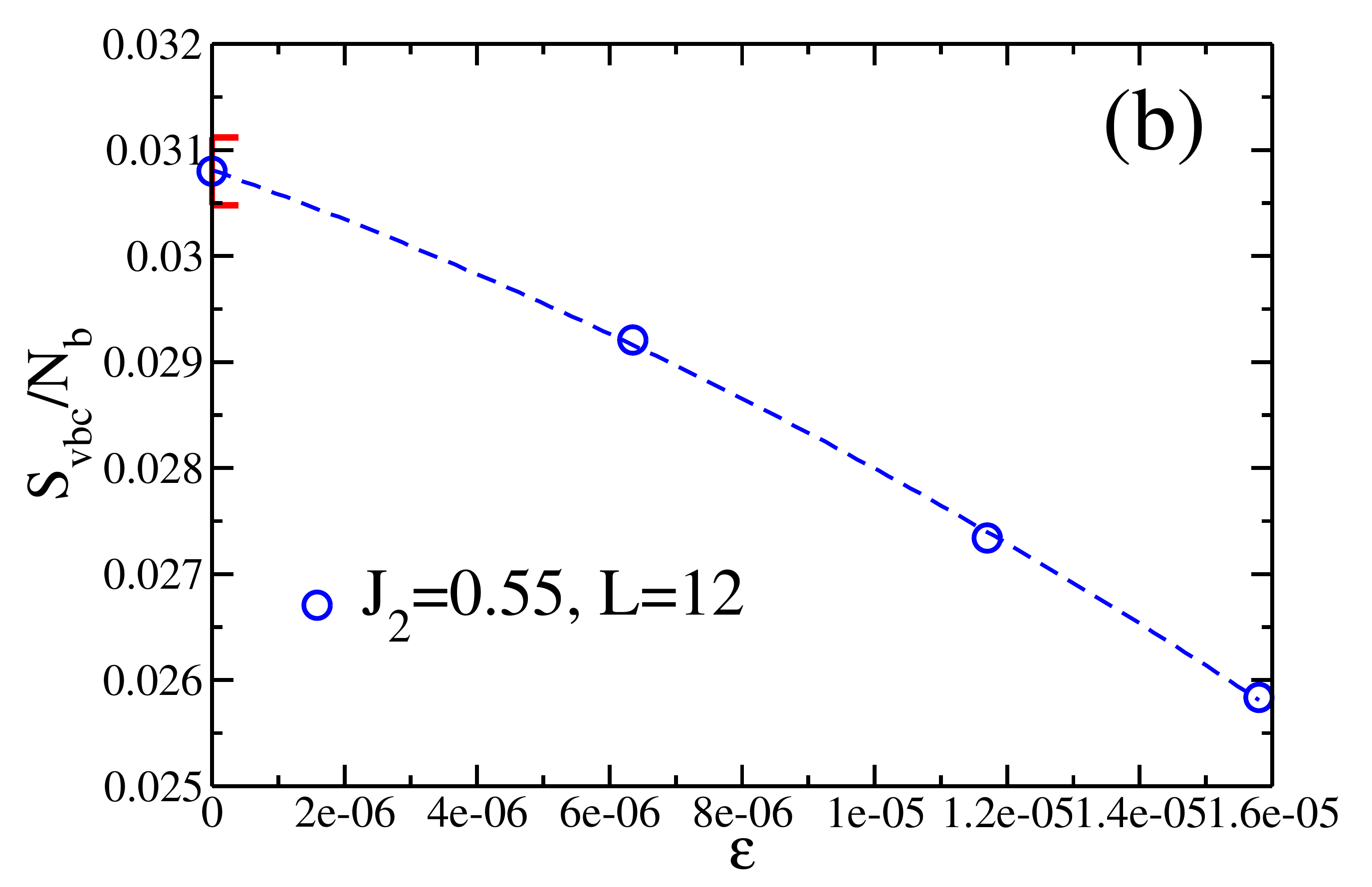}
\caption{(color online) Dimer structure factor $S_{\rm vbc}/N_b$ versus the DMRG truncation error $\varepsilon$ for $J_2 = 0.55$ on (a) RC8-16 and (b) RC12-24 cylinders. For the RC12-24 cylinder at $J_2 = 0.55$, we reach the truncation error $6\times 10^{-6}$ by keeping up to $20000$ $U(1)$ equivalent states, and we extrapolate the data to estimate the result for $\varepsilon = 0$ using the extrapolation function $S_{\rm vbc}(\varepsilon)/N_b=S_{\rm vbc}(0)/N_b+a\varepsilon+b\varepsilon^2$. The red lines indicate the extrapolation error bar.}\label{dimer_scaling}
\end{figure}

\begin{figure}[tbp]
\includegraphics[width = 1.0\linewidth,clip]{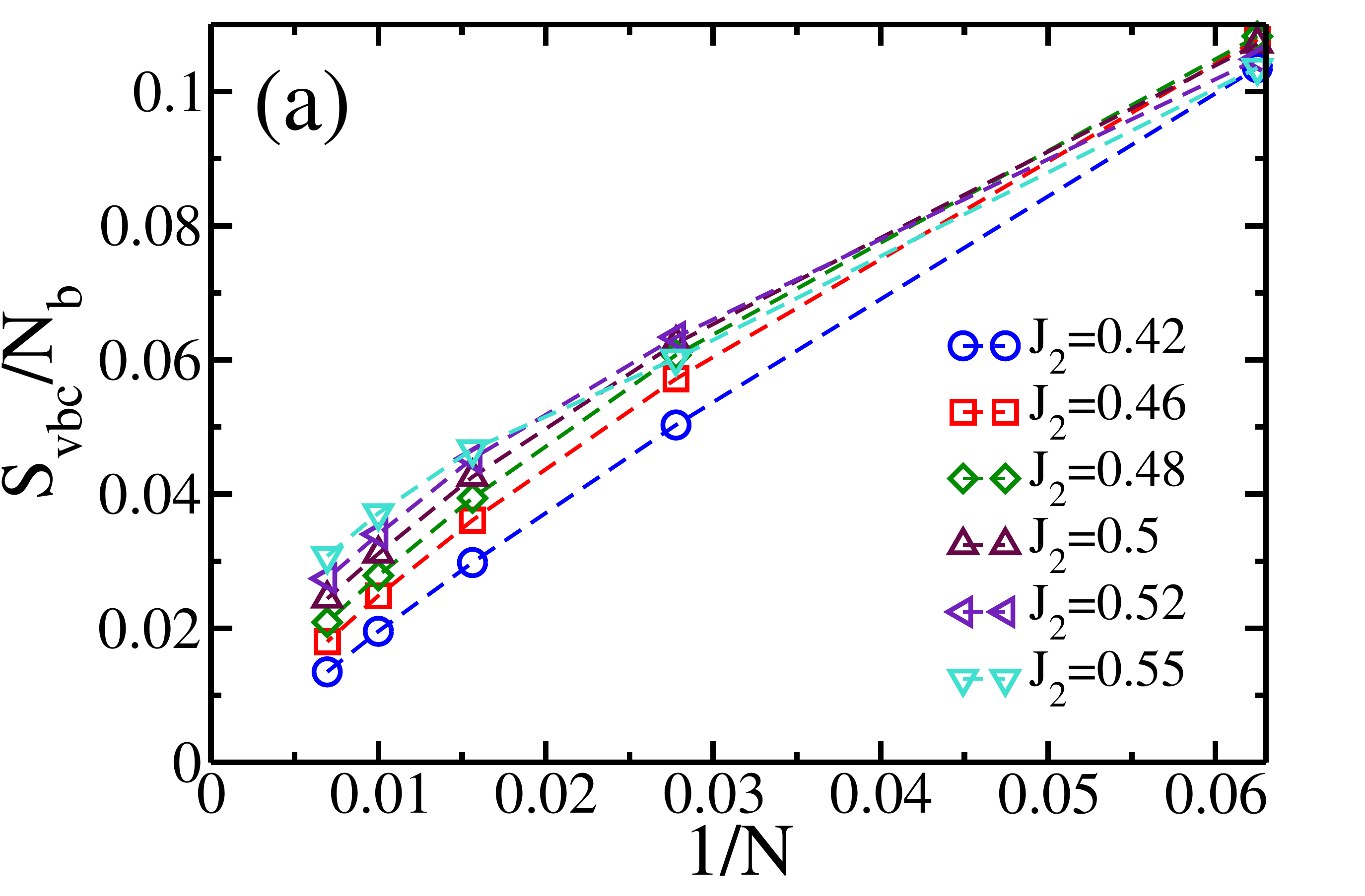}
\includegraphics[width = 1.0\linewidth,clip]{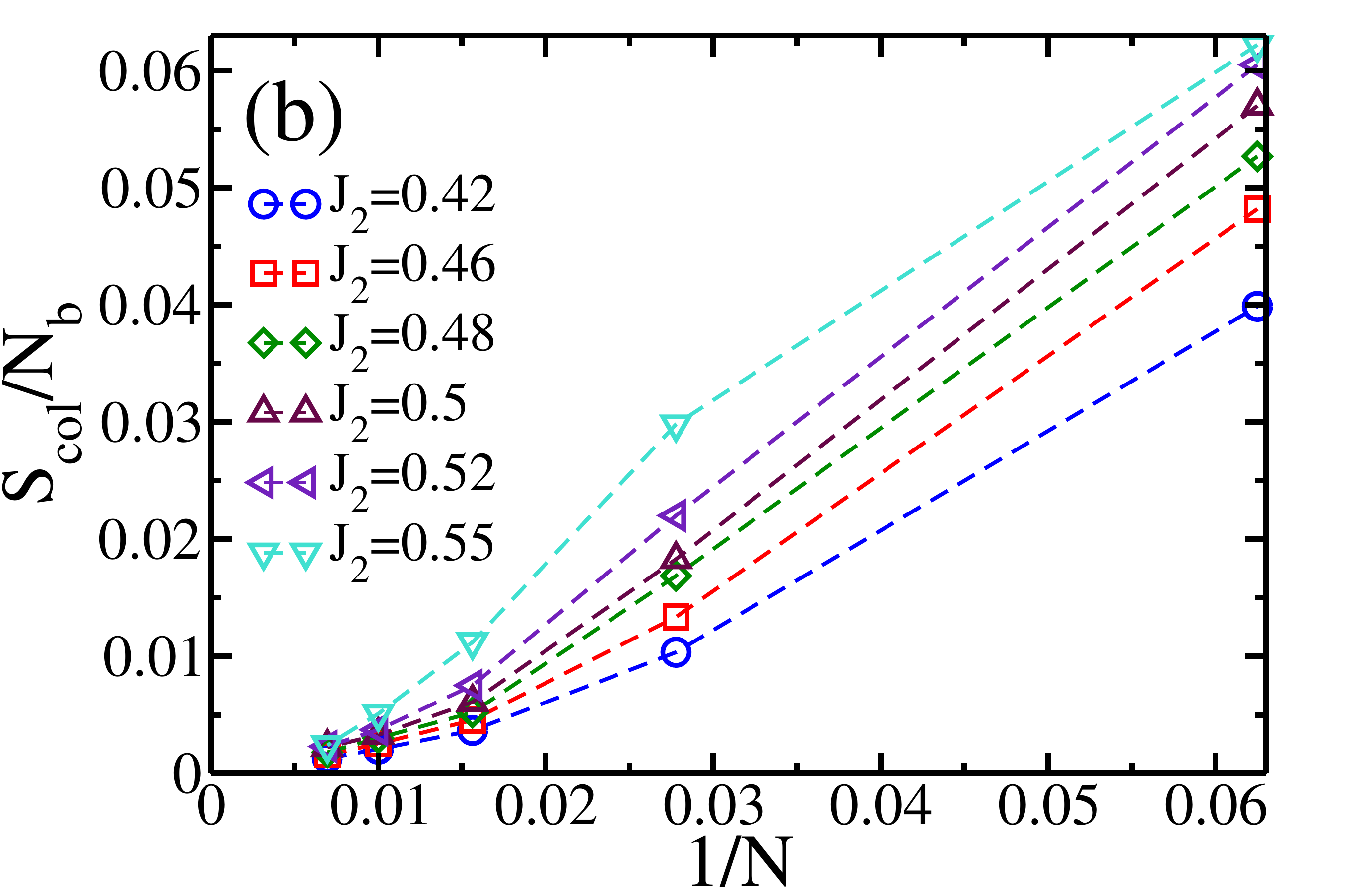}
\includegraphics[width = 1.0\linewidth,clip]{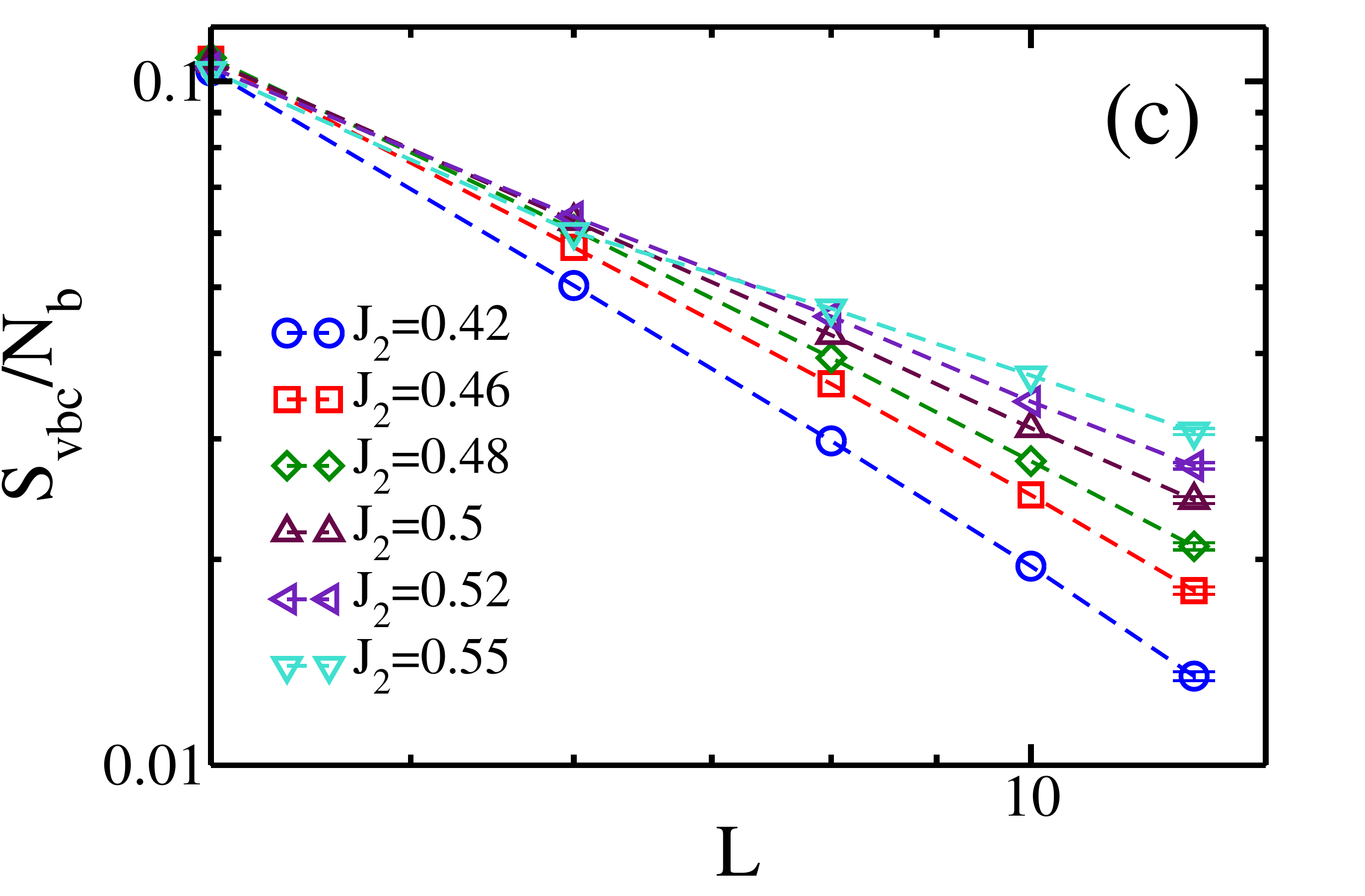}
\caption{(color online) Size dependence of dimer structure factors (a) $S_{\rm vbc}/N_b$ and (b) $S_{\rm col}/N_b$.
$S_{\rm vbc}/N_b$ appears to extrapolate to finite values for $J_2 > 0.5$, while $S_{\rm col}/N_b$ decays quite fast
and approaches zero with increasing system size thus excluding the CVB order.
(c) Log-log plot of $S_{\rm vbc}/N_b$ versus width $L$. The error bars for $L = 12$ data denote the extrapolation uncertainties with DMRG truncation error as shown in Fig.~\ref{dimer_scaling}(b).}\label{dimer_factor}
\end{figure}

\textit{Dimer structure factors on RC cylinder.---} The CVB order breaks rotational symmetry, while the PVB order preserves it.  Following Ref.~\cite{PRB_74_144422}, we consider two structure factors $S_{\rm vbc}$ and $S_{\rm col}$ obtained from the dimer-dimer correlations. $S_{\rm vbc}$ diverges in both the CVB and PVB states, while $S_{\rm col}$ diverges only in the CVB state.  The structure factors are defined as
\begin{equation}
S_{\lambda} = \sum_{(k,l)}\varepsilon_{\lambda}(k,l)C_{ijkl}, \label{dimer_stru}
\end{equation}
where $\lambda$ is either ``vbc" or ``col". The phase factors $\varepsilon_{\lambda}(k,l)$ are shown in Fig.~\ref{definition}, which reproduces Fig.~7 in Ref.~\cite{PRB_74_144422}. Dimer-dimer correlation function $C_{ijkl}$ is defined as
\begin{equation}
C_{ijkl} = 4\left[\langle (S_i \cdot S_j)(S_k \cdot S_l)\rangle - \langle S_i \cdot S_j\rangle\langle S_k \cdot S_l\rangle \right].
\end{equation}
We calculate dimer-dimer correlation function on the RC$L$-$2L$ cylinder with a reference bond $(i,j)$ in the middle of the cylinder (we have considered both horizontal and vertical reference bonds but will show only the former).  Figure~\ref{dimer_pattern} shows the dimer-dimer correlations on the RC10-20 cylinder at $J_2 = 0.55$ with the reference bond ($i,j$) oriented horizontally in the middle of the cylinder.  The red and blue bonds indicate negative and positive dimer correlations, respectively.  We see alternating red and blue horizontal bonds of comparable strengths, while the vertical bonds show significantly weaker correlations; this picture looks much more like the pattern of the pure $s$-wave plaquette state (PVB) in Table~III of Ref.~\cite{PRB_74_144422} rather than the pattern of the pure columnar state.

Figures~\ref{dimer_scaling}~and~\ref{dimer_factor} shows our best estimates of the
structure factors $S_{\rm vbc}/N_b$ and $S_{\rm col}/N_b$ obtained with a horizontal reference bond ($i,j$) and normalized by the number of bonds $N_b$ used to calculate the structure factors.
We note that we need to pay much attention to the DMRG convergence to obtain reliable results.  Figure~\ref{dimer_scaling} shows the structure factor $S_{\rm vbc}/N_b$ versus the DMRG truncation error $\varepsilon$ at $J_2 = 0.55$.  From Fig.~\ref{dimer_scaling}(a), we can obtain accurate results (within $0.5\%$) for the RC8-16 cylinder by just using the data with the smallest $\varepsilon$. We can similarly obtain accurate results for the RC10-20 cylinder (not shown) with $\varepsilon \simeq 1\times 10^{-6}$. However, for the RC12-24 cylinder, our truncation error is still $\varepsilon \simeq 6\times 10^{-6}$ even when we keep as many as $20000$ $U(1)$ equivalent states.  In this case, we extrapolate the data with $\varepsilon$ to estimate the result for $\varepsilon = 0$ as shown in Fig.~\ref{dimer_scaling}(b). The extrapolation function is $S_{\rm vbc}(\varepsilon)/N_b=S_{\rm vbc}(0)/N_b+a\varepsilon+b\varepsilon^2$. Without such an extrapolation, we would underestimate the magnitude of the VBC order parameter by about $7\%$.  The error bar in Fig.~\ref{dimer_scaling}(b) is from the extrapolation uncertainty.

In Fig.~\ref{dimer_factor}(a), we see that $S_{\rm vbc}/N_b$ approaches zero upon increasing system size for $J_2 < 0.5$ and possibly extrapolates to finite values for $J_2 > 0.5$ if we fit the
large-size data using polynomials of $1/N$.
This suggests PVB or CVB orders at $J_2  > 0.5$. 
In Fig.~\ref{dimer_factor}(b), we see that $S_{\rm col}/N_b$ decays quite fast with system size and always approaches zero in the thermodynamic limit, which implies vanishing CVB order. Thus, the behavior of these two structure factors reveals the possible PVB order at $J_2 > 0.5$ and clearly excludes the CVB order.  We observe similar results with a vertical reference bond ($i,j$) (not shown).

We also notice that when we plot $S_{\rm vbc}/N_b$ versus $1/L$ ($L = \sqrt{N}$), the data could be extrapolated to zero or small values also for $J_2 > 0.5$, which would be similar to the analysis in Ref.~\cite{PRB_86_024424}.
However, for $J_2 = 0.55$, the extrapolation function to zero is almost linear in $1/L$ (plot not shown), while in a phase with no VBC order we would expect $S_{\rm vbc}/N_b$ to vanish as $1/N \sim 1/L^2$.  Thus this data is not consistent with vanishing VBC order.

In Fig.~\ref{dimer_factor}(c), we show log-log plot of $S_{\rm vbc}/N_b$ versus cylinder width $L$, which provides more insight about the transition region.  For $J_2 < 0.5$, the accelerated decay of $S_{\rm vbc}/N_b$ is consistent with vanishing dimer order.  The  finite-size data at $J_2 = 0.5$ and $0.52$ appear close to critical, with power law behavior $S_{\rm vbc}/N_b \sim L^{-1.4}$.  On the other hand, at $J_2 = 0.55$ the $S_{\rm vbc}/N_b$ decays significantly more slowly, which is consistent with a VBC order.  Overall, our structure factor results are consistent with the phase diagram in the main text obtained from the studies of the VBC texture decay lengths.

\textit{Dimer order of the next-nearest-neighbor bonds.---}We also studied the dimer order of the next-nearest-neighbor (NNN) $J_2$
bonds by investigating the NNN bond energy textures and how their decay length depends on the cylinder width.

\begin{figure}[tbp]
\includegraphics[width = 1.0\linewidth,clip]{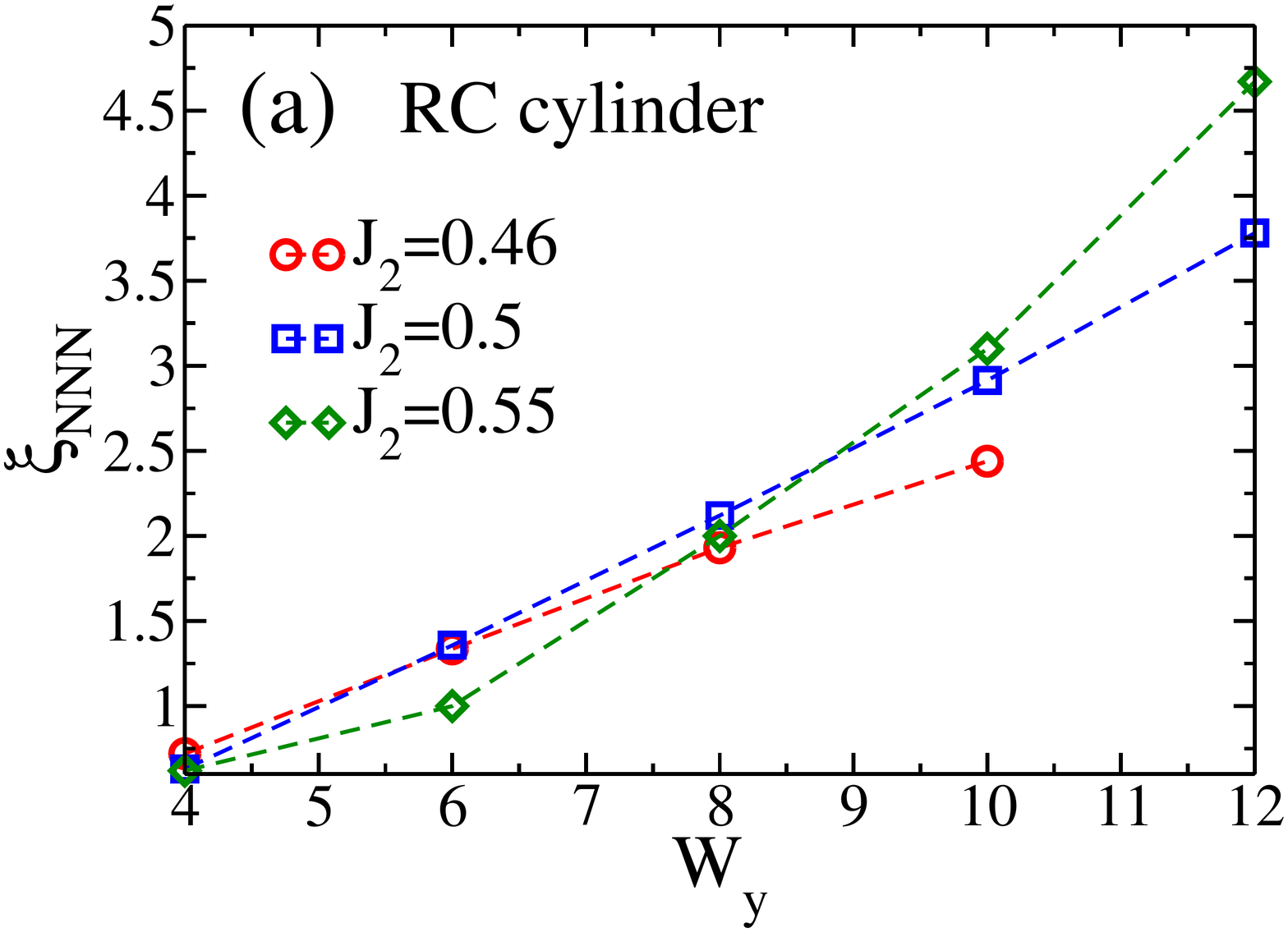}
\includegraphics[width = 1.0\linewidth,clip]{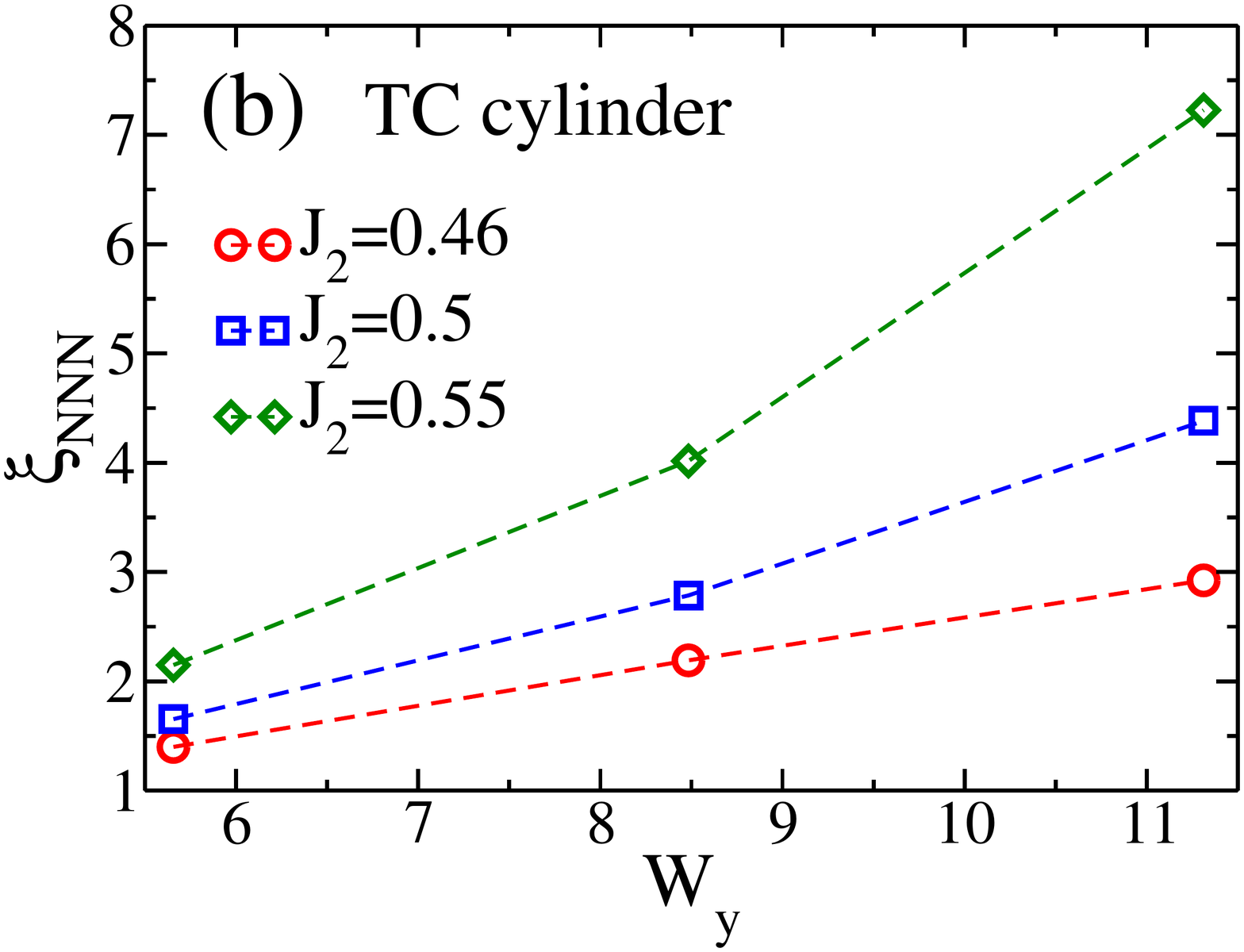}
\caption{(color online) (a) Dependence of the decay length of the NNN dimer order $\xi_{\rm NNN}$ with system width $W_y$ on the RC cylinder without boundary pinning.
(b) Dependence of $\xi_{\rm NNN}$ with system width $W_y$ on the TC cylinder.  On this cylinder, we find that both the horizontal and vertical dimer orders have the same $\xi_{\rm NNN}$.
On both cylinders, $\xi_{\rm NNN}$ grows faster than linearly for $J_2 > 0.5$ and saturates for $J_2 < 0.5$, consistent with the behavior of the decay length of the NN dimer orders.} \label{NNN}
\end{figure}

On the RC cylinder without additional boundary pinning, the diagonal NNN bond energy is the same inside each column but depends on
the column distance from the boundary. We define the NNN dimer order parameter as the difference of the NNN bond energies in
adjacent columns. By studying the NNN dimer order parameter on long cylinders, we find that it decays exponentially from the
boundary to the bulk with a decay length $\xi_{\rm NNN}$. As shown in Fig.~\ref{NNN}(a), $\xi_{\rm NNN}$ grows faster than linearly
for $J_2 > 0.5$, while it saturates for $J_2 < 0.5$; these results are consistent with the behavior of the decay length of the NN
bond texture. 

On the TC cylinder, the NNN bonds are either horizontal or vertical. We find that in both cases the textures have the same decay
length, which exhibits the same behavior with increasing system width as the NN bond decay length, see Fig.~\ref{NNN}(b).
Thus, similar to the NN bonds, the NNN bonds also appear to have a PVB order in the 2D limit.  Such results for the NNN bonds are
expected and do not represent a new VBC state but generally confirm the VBC order established from the NN bond energy studies.
Indeed, in the PVB phase, we expect the NNN bonds on the strong plaquettes to have somewhat different bond energy than the NNN
bonds on the weak plaquettes, and our results are consistent with such expectations.

\textit{Comparisons of torus energies from DMRG and VMC.---}We have compared our DMRG ground-state energies on $L \times L$ tori to VMC results with additional Lanczos improvement steps from Ref.~\cite{PRB_88_060402}.
Since the torus system is extremely difficult to fully converge for $8 \times 8$ or larger sizes, we keep up to $32000$ states and 
extrapolate the energy with the DMRG truncation error\cite{converge}. The extrapolated DMRG and Lanczos-VMC results are quite close to each other in the possible SL region $0.45<J_2<0.5$, indicating that the gapless $Z_2$ SL of Ref.~\cite{PRB_88_060402} has very competitive energies in this region.

\begin{figure}[tbp]
\includegraphics[width = 1.0\linewidth,clip]{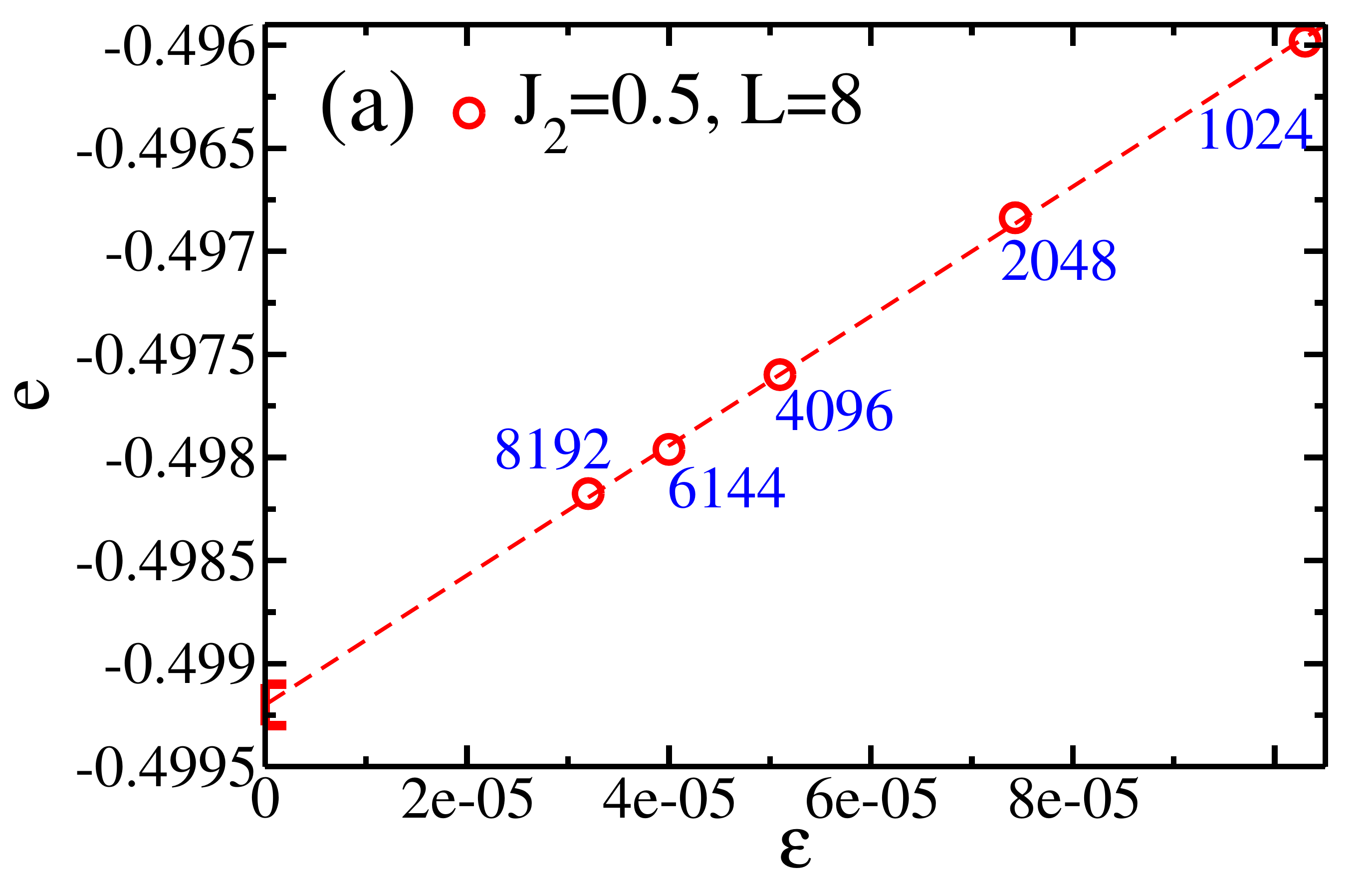}
\includegraphics[width = 1.0\linewidth,clip]{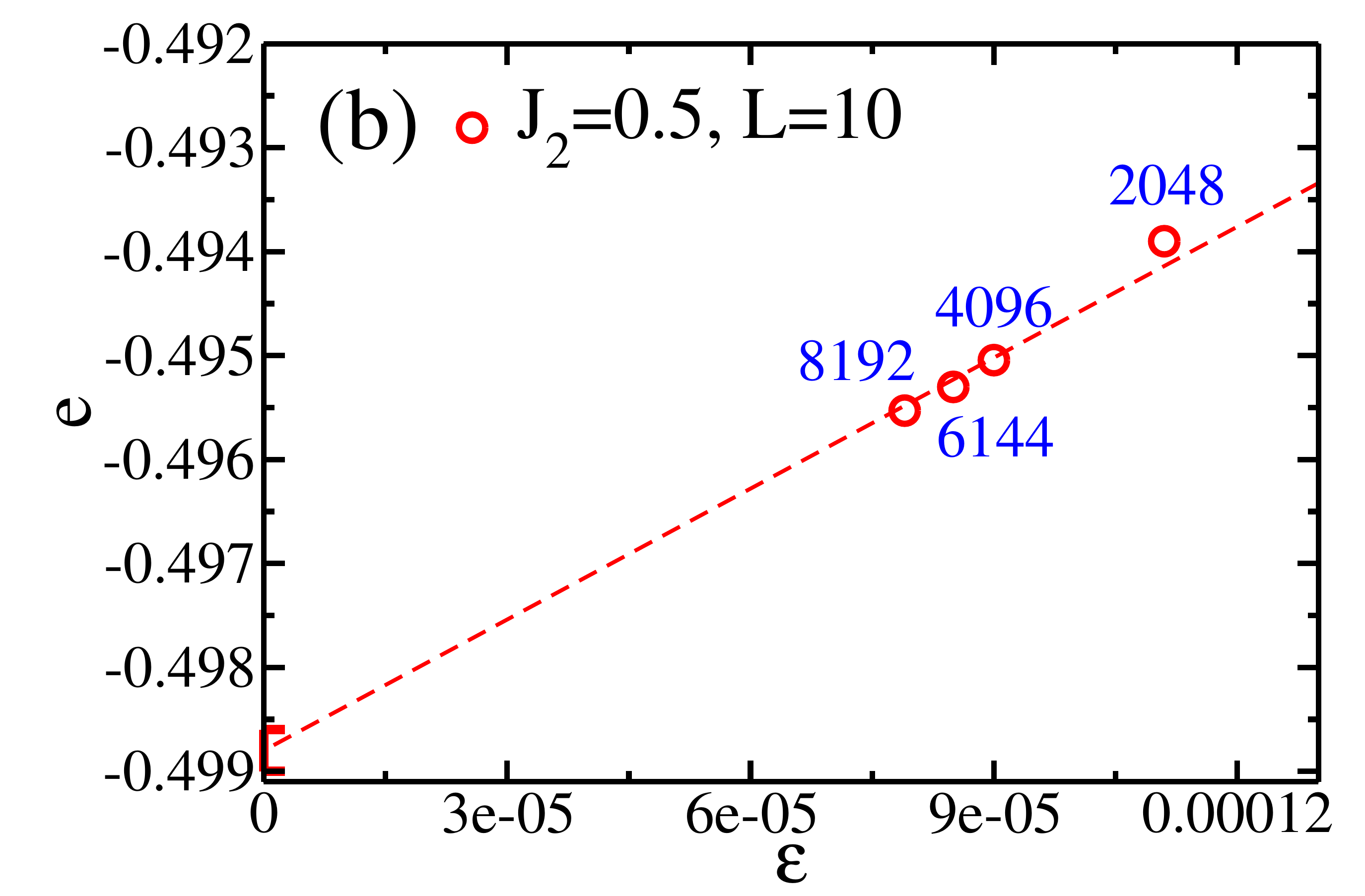}
\caption{(color online) Ground-state energy per site $e$ versus DMRG truncation error $\varepsilon$ on the $L \times L$ torus systems for (a) $J_2 = 0.5$, $L = 8$, and (b) $J_2 = 0.5$, $L = 10$. The numbers in the figures denote the kept $SU(2)$ states $M$ for obtaining the energy. We extrapolate the data with a straight line fitting and denote the $\varepsilon \to 0$ intercept (corresponding to $M \to \infty$) as DMRG ($\infty$).} \label{DMRG_energy}
\end{figure}

Table~\ref{energy_table} shows energy comparisons of DMRG and VMC for $J_2 = 0.4,0.45,0.5$, and $0.55$; it includes DMRG results obtained by keeping $4096$, $6144$, and $8192$ $SU(2)$ states [equivalent to about $16000$, $24000$, and $32000$ $U(1)$ optimal states], as well as VMC results with Lanczos improvement steps from Ref.~\cite{PRB_88_060402}. DMRG ($\infty$) denotes the DMRG energy extrapolated with truncation error;
as illustrated in Fig.~\ref{DMRG_energy}, we extrapolate the data points using a straight line fitting. VMC ($\rm p$=$\infty$) denotes the VMC energy extrapolations with the variance in Ref.~\cite{PRB_88_060402}.
The overall agreement shows, on one hand, that the DMRG is performing reasonably well even in the most challenging torus geometry.  Here we emphasize that all results in the main text are obtained using cylinder geometry where the DMRG measurements are much better converged\cite{converge} and represent essentially exact unbiased measurements.  On the other hand, the excellent performance of the Lanczos-VMC method is also notable.  It would be interesting to see this method tried in the cylinder geometries and results subjected to the finite-size scaling analysis as in the present work.

\begin{table*}

\begin{tabular}{|c|c|c|c|c|c|c|c|c|}
\hline
$J_2=0.40$ & DMRG ($4096$) & DMRG ($6144$) & DMRG ($8192$) & DMRG ($\infty$) & VMC ($p=0$) & VMC ($p=1$) & VMC ($p=2$) & VMC ($p=\infty$)\\
\hline
$L=6$ & $-0.529734$  & $-0.529742$ & $-0.529744$ & $-0.529747(1)$ & $-0.52715(1)$ & $-0.52928(1)$ & $-0.52957(1)$ & $-0.52972(1)$ \\
\hline
$L=8$ & $-0.524648$ & $-0.525013$ & $-0.525196$ & $-0.5262(1)$ & $-0.52302(1)$ & $-0.52501(1)$ & $-0.52539(1)$ & $-0.52556(1)$\\
\hline
$L=10$ & $-0.521487$ & $-0.522043$ & $-0.522391$ & $-0.5253$ & $-0.52188(1)$ & $-0.52368(1)$ & $-0.5240(1)$ & $-0.52429(2)$  \\
\hline
\hline
$J_2=0.45$ & DMRG ($4096$) & DMRG ($6144$) & DMRG ($8192$) & DMRG ($\infty$) & VMC ($p=0$) & VMC ($p=1$) & VMC ($p=2$) & VMC ($p=\infty$)\\
\hline
$L=6$ & $-0.515637$  & $-0.515652$ & $-0.515655$ & $-0.515660(1)$ & $-0.51364(1)$ & $-0.51538(1)$ & $-0.51558(1)$ & $-0.51566(1)$ \\
\hline
$L=8$ & $-0.510162$ & $-0.510534$ & $-0.510740$ & $-0.5116(1)$ & $-0.50930(1)$ & $-0.51101(1)$ & $-0.51125(1)$ & $-0.51140(1)$\\
\hline
$L=10$ & $-0.507193$ & $-0.507677$ & $-0.507976$ & $-0.5110$ & $-0.50811(1)$ & $-0.50973(1)$ & $-0.51001(1)$ & $-0.51017(2)$  \\
\hline
\hline
$J_2=0.50$ & DMRG ($4096$) & DMRG ($6144$) & DMRG ($8192$) & DMRG ($\infty$) & VMC ($p=0$) & VMC ($p=1$) & VMC ($p=2$) & VMC ($p=\infty$)\\
\hline
$L=6$ & $-0.503771$  & $-0.503797$ & $-0.503805$ & $-0.503808(1)$ & $-0.50117(1)$ & $-0.50323(1)$ & $-0.50357(1)$ & $-0.50382(1)$ \\
\hline
$L=8$ & $-0.497598$ & $-0.497961$ & $-0.498175$ & $-0.4992(1)$ & $-0.49656(1)$ & $-0.49855(1)$ & $-0.49886(1)$ & $-0.49906(1)$\\
\hline
$L=10$ & $-0.495044$ & $-0.495301$ & $-0.495530$ & $-0.4988$ & $-0.49521(1)$ & $-0.49718(1)$ & $-0.49755(1)$ & $-0.49781(2)$  \\
\hline
\hline
$J_2=0.55$ & DMRG ($4096$) & DMRG ($6144$) & DMRG ($8192$) & DMRG ($\infty$) & VMC ($p=0$) & VMC ($p=1$) & VMC ($p=2$) & VMC ($p=\infty$)\\
\hline
$L=6$ & $-0.495096$ & $-0.495150$ & $-0.495167$ & $-0.495186(1)$ & $-0.48992(1)$ & $-0.49303(1)$ & $-0.49399(1)$ & $-0.49521(7)$\\
\hline
$L=8$ & $-0.487685$ & $-0.487982$ & $-0.488160$ & $-0.4891(1)$ & $-0.48487(1)$ & $-0.48777(1)$ & $-0.48841(2)$ & $-0.48894(3)$\\
\hline
$L=10$ & $-0.484890$ & $-0.485239$ & $-0.485434$ & $-0.4880$ & $-0.48335(1)$ & $-0.48622(1)$ & $-0.48693(3)$ & $-0.48766(6)$  \\
\hline
\end{tabular}
\caption{DMRG and VMC ground-state energies on $L\times L$ tori with $J_2=0.4,0.45,0.5$ and $0.55$. DMRG energies are obtained by keeping $4096$, $6144$, and $8192$ $SU(2)$ states.
DMRG ($\infty$) is obtained from the straight line energy extrapolation with DMRG truncation error as illustrated in Fig.~\ref{DMRG_energy}.
For $L = 10$, we do not show the error bar of extrapolation because the DMRG truncation error is large ($\epsilon\simeq 8\times 10^{-5}$) and thus our estimation of the error bar is not accurate. The extrapolated results for $L = 10$ are obtained from the linear fitting of the data by keeping $4096$, $6144$ and $8192$ $SU(2)$ states.
The VMC energies are from Ref.~\cite{PRB_88_060402}; $p$ denotes the Lanczos step; and VMC ($p=\infty$) is obtained from extrapolation with the variance.}\label{energy_table}
\end{table*}

\begin{figure}[tbp]
\includegraphics[width = 1.0\linewidth,clip]{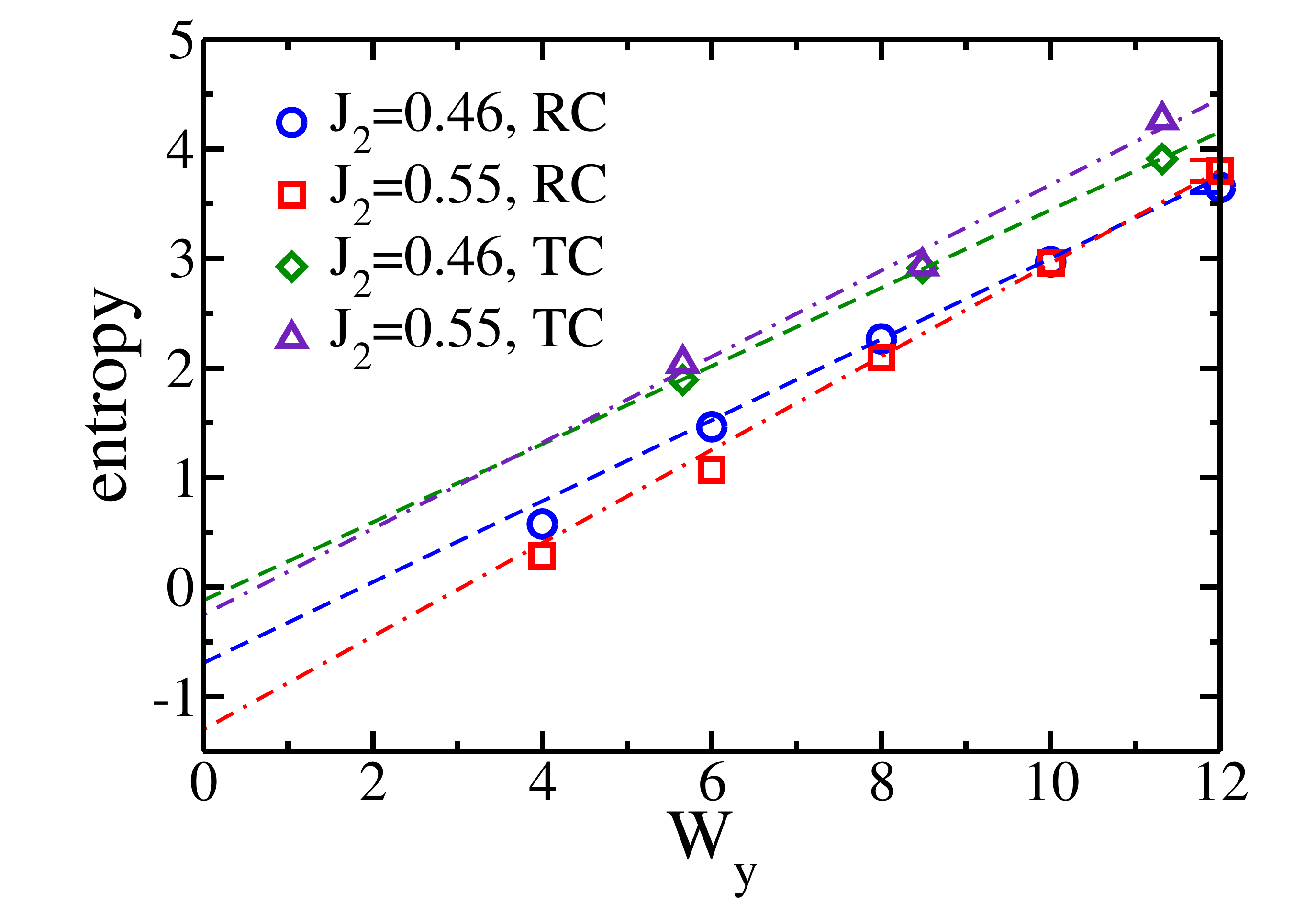}
\caption{(color online) Entanglement entropy as a function of system width on RC and TC cylinders. For each width, we obtain the entropy by extrapolating measurements on long cylinders to $L_x \rightarrow \infty$ limit.} \label{entropy}
\end{figure}

\textit{Entanglement entropy.---}For gapped quantum states with topological order, topological entanglement entropy (TEE) $\gamma$ is proposed to characterize non-local feature of entanglement\cite{PRL_96_110404,PRL_96_110405}. The Renyi entropies of a subsystem $A$ with reduced density matrix $\rho_A$ are defined as $S_n = (1-n)^{-1}\ln({\rm Tr}\rho_A^{n})$;
$n \rightarrow 1$ limit gives the Von Neuman entropy.  For a topologically ordered state, Renyi entropies have the form $S_n = \alpha L - \gamma$, where $L$ is the boundary of the subsystem, and all other terms vanish in the large $L$ limit; $\alpha$ is a non-universal constant, while a positive $\gamma$ is a correction to the area law of entanglement and reaches a universal value determined by total quantum dimension $D$ of quasiparticle excitations\cite{PRL_96_110404,PRL_96_110405}.
Previous DMRG study\cite{PRB_86_024424} found $\gamma \approx \ln 2$ in the intermediate region of $J_2$ consistent with a $Z_2$ SL for this model. We compute the entanglement entropy (EE) on long cylinders by partitioning the system in the middle along the vertical direction. For each fixed $L_y$, we fit the entropy to $L_x \rightarrow \infty$ limit to find the entropy of a possible minimum entropy state\cite{NP_8_902}.

In Fig.~\ref{entropy}, we show our DMRG results for the EE at $J_2 = 0.46$ and $0.55$ on both TC and RC cylinders. We obtain accurate EE when $W_y < 12$. For $W_y = 12$, we extrapolate the EE with the DMRG truncation error, which has significant uncertainty from the extrapolation.
On RC cylinder, we perform linear fit of the EE versus $W_y$ using the three largest sizes.
We find the TEE at $J_2 = 0.46$ is close to $\ln 2$, while at $J_2 = 0.55$ is close to $-1.3$. However, the system appears to have large finite-size effects, which can be seen by comparing the results on the RC and TC cylinders. On the TC cylinder, the linear fits of the EE vs $W_y$ give the TEE close to zero, which is different from the RC cylinder. 
Similar effect has also been observed in the $J_1$-$J_2$ model on the honeycomb lattice\cite{PRL_110_127205, arxiv_1306_6067}. Because of such strong finite-size effects, the TEE obtained by fitting EE on our small sizes may not be able to distinguish different quantum phases in the $J_1$-$J_2$ square lattice model.

\end{appendices}


\begin{thebibliography}{99}
\bibitem{converge} We get much better convergence on cylinder than on torus. We can achieve the
truncation error $1 \times 10^{-6}$ for $L_y = 10$ cylinder and $5\times 10^{-6}$ for $L_y = 12$ cylinder. For example, for $J_2 = 0.5$ and $L_y = 10$, we get the truncation error $1 \times 10^{-6}$ on cylinder by keeping $20000$ states, while the error is much larger $8 \times 10^{-5}$ on $10 \times 10$ torus even when we keep $32000$ states.
\bibitem{Neel} We get converged $m_{s}^{2}$ for $L \leq 12$ by keeping more than $20000$ states, while the results for $L = 14$ are obtained through extrapolations with the DMRG truncation error.
\bibitem{discrepancy} In $U(1)$ DMRG calculations for $L_y \geq 10$, one needs to extrapolate $\xi_{\rm y}$ with the DMRG truncation error, which may give some uncertainty to $\xi_{\rm y}$.
1
\end{thebibliography}
\end{document}